\documentclass[twocolumn,twoside,trackchanges]{aastex701}
\usepackage{amsmath}

\usepackage{subcaption}
\usepackage{soul}
\usepackage{amssymb,microtype,siunitx,booktabs}

\begin{document}

\title{Kinematic and Dynamical Properties of Solar Neighborhood White Dwarfs from SDSS DR19 and Gaia DR3}

\correspondingauthor{Ozcan Caliskan}

\author[0009-0003-5839-8007]{Ozcan Caliskan}
\affiliation{Programme of Astronomy and Space Sciences, Institute of Graduate Studies in Science, Istanbul University, 34116, Istanbul, T\"{u}rkiye}
\email[show]{caliskanozcan@hotmail.com}  

\author[0000-0003-3510-1509]{Sel\c{c}uk Bilir}
\affiliation{Department of Astronomy and Space Sciences, Faculty of Science, Istanbul University, 34119, Beyaz\i t, Istanbul, T\"{u}rkiye}
\email{sbilir@istanbul.edu.tr}

\author[0000-0001-6098-2235]{Mukremin Kilic}
\affiliation{Homer L. Dodge Department of Physics and Astronomy, University of Oklahoma, 440 W. Brooks Str., Norman, OK, 73019, USA}
\email{mukremin@gmail.com}

\begin{abstract}

We present a comprehensive kinematic and dynamical analysis of 5,837 DA white dwarfs with SDSS DR19 spectroscopy and Gaia DR3 astrometry. Using complementary spatial distributions, kinematic properties, and orbital parameters, we show that the sample is overwhelmingly dominated by the thin disk, with only minor contributions from the thick disk and the halo. Velocity dispersions and orbital eccentricities increase systematically from the thin disk to the halo, reflecting progressively hotter Galactic populations. Total stellar ages available for 4,829 white dwarfs reveal a clear age--velocity dispersion relation with power-law indices of 0.06--0.17, confirming that white dwarfs preserve the secular dynamical heating history of the Galactic disk. We derive a thin-disk scale-height of 123--229 pc and radial scale-length of 3.2--3.5 kpc, while the thick disk has a scale height of 865 pc and contributes approximately 6--7\% of the local stellar density. Reconstructed birth radii indicate that changes in the guiding radii represent an important component of the present-day radial redistribution of nearby white dwarfs, although the relative contributions of radial migration and orbital blurring cannot be uniquely separated from these displacement measures. These results establish white dwarfs as powerful tracers of the kinematic, dynamical, and structural evolution of the Milky Way.

\end{abstract}




\section{Introduction}
\label{sec:intro}

White dwarfs (WDs) constitute the final evolutionary products of nearly all low- and intermediate-mass stars and are therefore among the most abundant stellar remnants in the Milky Way. Their predictable cooling evolution \citep{Mestel52} and relatively simple internal structure make them fundamental laboratories for studying stellar evolution and the Galactic structure \citep{Fontaine2001, Althaus2010}. Because WDs cool over Gyr timescales, their present-day kinematic properties retain important signatures of the dynamical heating, radial migration, and secular evolution processes that have shaped the Galactic disk throughout its history \citep{Winget1987, Liebert1988, Tremblay2014}. Their kinematics and Galactic orbits provide valuable constraints on the age distribution, dynamical evolution, and the structure of the Milky Way disk and halo \citep{Pauli2006, Anguiano2017, Kilic2017, Raddi2022}.

Over the last quarter century, large-scale photometric, astrometric and spectroscopic surveys, particularly the Sloan Digital Sky Survey \citep[SDSS,][]{York2000} and the Gaia mission \citep{Gaia2018}, have revealed that the Galactic disk exhibits a rich and dynamically complex structure shaped by processes such as secular heating, radial migration, spiral arm perturbations, bar resonances, and interactions with satellite galaxies \citep[e.g.,][]{Sellwood2002, Schonrich2009, Minchev2010, Antoja2018, Helmi2018}. WDs constitute particularly valuable kinematic tracers as they represent stellar populations spanning a wide range of ages \citep{Pauli2006, GentileFusillo2019, JimenezEsteban2018}. Specifically, the observed relation between the stellar age and the space velocity dispersion provides one of the strongest empirical constraints on the models of Galactic disk heating and dynamical evolution \citep{Wielen1977, Aumer2009, Ting2019}. \citet{Cheng2019} used a photometrically selected sample of high-mass WDs from Gaia Data Release 2 \citep[Gaia DR2;][]{Gaia2018} to model their transverse-velocity distributions as a function of age, and demonstrate the value of WD kinematics as an independent constraint on stellar ages.

Before Gaia, kinematic investigations of WD populations were generally restricted to relatively small and heterogeneous samples due to the limited availability of accurate trigonometric parallaxes, proper motions, and radial velocities. Significant progress was made with the ESO Supernovae Type Ia Progenitor Survey (SPY), which provided high-resolution spectroscopic observations and accurate radial velocity measurements for several hundred DA WDs \citep[e.g.,][]{Napiwotzki2001, Pauli2003, Karl2005}. By combining spectroscopic radial velocities with proper motions and spectroscopic parallax estimates, these studies enabled a more reliable separation of thin-disk, thick-disk, and halo populations based on their Galactic space motions and orbital parameters \citep{Pauli2006}. 

\citet{Anguiano2017} significantly expanded upon this work by using the DA white dwarf sample from the SDSS DR12; they confirmed the existence
of a mean Galactocentric radial velocity gradient, and found evidence of an additional source of heating in the disk. The SPY and the SDSS DR12 samples highlighted the importance of full three-dimensional kinematic information for distinguishing Galactic populations and constraining the evolutionary history of WDs in the Solar neighborhood. Nevertheless, pre-Gaia investigations remained limited by small sample sizes, selection biases, or incomplete phase-space information. 

\citet{Torres2019} applied a random forest algorithm trained on population-synthesis models to distinguish thin-disk, thick-disk, and halo stars in the Gaia DR2 100 pc WD sample. Their analysis demonstrated that probabilistic classification based on astrometric and photometric information provides a complementary approach to conventional kinematic criteria for separating Galactic WD populations.

Recently, \citet{Raddi2022} took advantage of the Gaia Early Data Release 3 \citep[Gaia EDR3;][]{GaiaEDR3} astrometry to revisit the SPY and SDSS DR12 DA white dwarf samples. They also included common proper motion pairs of white dwarfs with non-degenerate companions, where the radial velocity or metallicity of the companion is constrained from spectroscopy. \citet{Raddi2022} found that their sample is dominated (90-95\%) by thin disk stars, with negligible contribution from the halo. Studying the age-velocity dispersion relation (AVR) for the disk, they found evidence of dynamical heating and saturation after 4-6 Gyr, which could be driven by giant molecular clouds, the spiral structure, Galactic bar perturbations, and merger events \citep{Lacey1984, DeSimone2004, Grand2016}. However, the relative contributions of these processes remain a matter of active debate. More recently, \citet{Swan2026} presented a kinematic analysis of the Dark Energy Spectroscopic Instrument Data Release 1 (DESI DR1) WD sample \citep[see also,][]{Kilic26a, Kilic26b, Amorim26} combining spectroscopic radial velocities and distance estimates with Gaia proper motions. Using a Gaussian-mixture model applied to the three-dimensional velocity distributions, they found that only about 1\% of the stars have a probability $>50$\% of being a halo star, indicating that the DESI sample is also dominated by the Galactic disk.

The SDSS white dwarf sample has grown significantly since DR12 \citep{Kepler19, Kepler21, Crumpler2025}. In this study, we investigate the kinematic properties of a sample of more than 5,800 WDs by combining Gaia astrometric data with the latest SDSS spectroscopic measurements. Using this statistically significant sample, we analyze the Galactic velocity distributions and examine the kinematic behavior of the WD population in the context of Galactic disk evolution. The paper is organized as follows. In Section~\ref{sec:data}, we describe the sample selection and data used in this work. Section~\ref{sec:method} presents the methods employed to derive the kinematic and Galactic orbital parameters. The results of the analyses of the spatial distribution, kinematics, and Galactic orbital parameters of the WDs are presented in Section~\ref{sec:result}, while Section~\ref{sec:summary-and-discussion} summarizes the main findings and discusses their implications. We conclude in Section \ref{sec:con}.

\section{Data}
\label{sec:data}

Our study is based on the DA WD catalog of \cite{Crumpler2025}, which provides a large and comprehensive sample of spectroscopically confirmed WDs. This dataset encompasses observations from both the 19th Data Release (DR19) of the ongoing SDSS-V survey (8,545 DAs) and previous generations of the Sloan Digital Sky Survey (19,257 DAs), resulting in a sample of 26,041 unique DA WDs, after accounting for an overlap of 1,761 sources observed across both SDSS-V DR19 and earlier SDSS generations. This sample offers a homogeneous spectroscopic dataset optimized for kinematic studies, delivering precise radial velocity measurements, spectroscopic effective temperatures ($T_{\rm eff}$), and surface gravities ($\log g$) for hydrogen atmosphere WDs. In addition, high-precision proper motions and trigonometric parallaxes from Gaia DR3, and geometric distance estimates from \cite{Bailer-Jones2021} enable a full six-dimensional phase-space characterization of this WD sample. 

\begin{table}[t]
\centering
\small
\caption{Quality cuts for our WD catalog.}
\label{tab:quality_cuts}
\begin{tabular}{ll}
\toprule
\textbf{Criterion} & \textbf{Explanation} \\
\midrule
$\sigma_d / d \leq 0.20$ & Distance uncertainty limit\\
$\sigma_{V_{\rm Rad}} \leq 15~\mathrm{km\,s^{-1}}$ & RV uncertainty limit \\
\texttt{binary\_flag} = 0 & Exclusion of binaries \\
$\mathrm{SNR} \geq 10$ & Minimum SNR \\
$\mathrm{RUWE} \leq 1.4$ & Gaia astrometric quality cut \\
\midrule
\multicolumn{2}{c}{Final sample: 5,894 WDs} \\
\bottomrule
\end{tabular}
\end{table}

We construct our kinematic sample by applying a series of quality cuts, given in Table~\ref{tab:quality_cuts}, to ensure reliable astrometric and spectroscopic measurements. We retain only sources with relative distance uncertainties below 20\%, radial velocity uncertainties smaller than 15 km s$^{-1}$, and signal-to-noise ratio (SNR) $\geq 10$ spectra. We further exclude binary candidates by requiring \texttt{binary\_flag} = 0 and imposing the Gaia astrometric quality cut RUWE $\leq 1.4$. These selection criteria reduce the sample to 5,894 WDs.

\section{Methods}
\label{sec:method}

In this section, we present a unified analysis framework covering WD kinematics, Galactic orbital properties, population classification, age estimation, and Galactic structure. First, space velocity components are computed from astrometric data and available radial velocities. Radial velocities of WDs require special treatment because the measured spectroscopic velocities include both the intrinsic stellar motion and the gravitational redshift produced by the strong gravitational potential at the stellar surface \citep[e.g.,][]{Falcon2010}. Therefore, to derive reliable 3D kinematics, the observed radial velocities were corrected for gravitational redshift before computing the Galactic space motions. 

The effective temperatures ($T_{\rm eff}$) and surface gravities ($\log g$) adopted from \citet{Crumpler2025} were used as input to \texttt{wdwarfdate} \citep{Kiman2022} to infer the WD masses. The corresponding stellar radii were then derived from the relation $R_{\rm WD}=\sqrt{(GM_{\rm WD}/g})$. Using these internally consistent mass and radius estimates, the gravitational-redshift contribution was calculated individually for each WD as $ V_{\rm gr}=\frac{G M_{\rm WD}}{R_{\rm WD}c} =0.635\,\frac{M_{\rm WD}/M_\odot}{R_{\rm WD}/R_\odot} \ {\rm km~s^{-1}},$ and was subtracted from the measured apparent spectroscopic radial velocity, such that $V_{\rm rad}=RV_{\rm obs}-V_{\rm gr}$. Thus, the masses and radii used in the gravitational-redshift correction were not adopted directly from \citet{Crumpler2025}; rather, the WD masses were inferred with \texttt{wdwarfdate} from the spectroscopic parameters of \citet{Crumpler2025}, and the radii were subsequently derived from the corresponding masses and surface gravities. The corrected radial velocities were then used to calculate the Galactic space-velocity components. From these, Galactic orbital parameters such as apogalactic and perigalactic distances, eccentricity, and maximum vertical height from the Galactic plane are derived. The sample is then classified into thin-disk, thick-disk, and halo populations based on probabilistic kinematic criteria and Galactic orbital parameters. Finally, the Jeans equations are applied to the full sample to derive the radial and vertical scale lengths of the Galactic disk.

\subsection{White Dwarf Age Determination}
Total stellar ages and progenitor masses were determined with the Bayesian framework implemented in the \texttt{wdwarfdate} package \citep{Kiman2022}. For our DA WD sample, we adopted the thick-H atmosphere cooling sequences of \citet{Bedard2020}, the MIST-based initial--final mass relation (IFMR) of \citet{Cummings2018}, and the MESA Isochrones and Stellar Tracks (MIST) evolution models \citep{Choi2016, Dotter2016} to determine the progenitor lifetimes. The progenitor models were evaluated at Solar metallicity ($[\mathrm{Fe/H}]=0$ dex) with no rotation ($v/v_{\rm crit}=0$). The measured $T_{\rm eff}$ and $\log g$, together with their uncertainties, were supplied to \texttt{wdwarfdate}, which evaluates their posterior distributions and derives the WD mass and cooling age ($t_{\rm cool}$), the progenitor mass and lifetime ($t_{\rm MS}$), and the total stellar age ($t_{\rm tot}=t_{\rm cool}+t_{\rm MS}$). The quoted values are the posterior medians, while the lower and upper uncertainties are the 16th and 84th percentiles, respectively. The Bayesian calculation also accounts for the intrinsic scatter in the adopted IFMR.

We adopted Solar-metallicity progenitor models uniformly for the full sample. The metallicities of the progenitor stars cannot be measured directly for isolated WDs, and assigning a mean metallicity according to kinematic population membership would introduce an additional model-dependent assumption. Nevertheless, progenitor metallicity affects both the pre-WD evolutionary lifetime and, to some extent, the initial--final mass relation \citep[e.g.,][]{Romero2015}. Therefore, the total ages derived here should be interpreted within the adopted Solar-metallicity framework, particularly for thick-disk and halo WDs whose progenitors are expected, on average, to be more metal-poor. This metallicity dependence represents an additional systematic uncertainty in the total ages that is not included in the quoted statistical uncertainties.

\subsection{Space Velocities}
We computed the three-dimensional Galactic space-velocity components ($U$, $V$, $W$) following the general matrix formalism of \citet{Johnson1987}, while adopting the updated ICRS-based equatorial-to-Galactic transformation used in the Gaia astrometric framework \citep{ESA1997,GaiaEDR3}. This analysis requires four input parameters: the equatorial coordinates ($\alpha,\delta$), proper-motion components ($\mu_{\alpha}\cos\delta,\mu_{\delta}$), geometric distances ($d$), and radial velocities ($V_{\rm rad}$). The uncertainties in the Galactic space-velocity components were calculated individually for each WD following the error-propagation formalism of \citet{Johnson1987} and the observational uncertainties in proper motion, geometric distance, and radial velocity. For the geometric distances adopted from \citet{Bailer-Jones2021}, the catalog-reported distance uncertainties were used directly, without any additional transformation or re-estimation of the distance posterior. The Gaia astrometric covariance terms and the possible asymmetry of the individual distance posteriors were not explicitly incorporated into this analytic error propagation. In this study, we adopt a left-handed Galactic coordinate system, where $U$ is positive toward the Galactic Center ($\ell=0^{\circ},b=0^{\circ}$), $V$ is positive in the direction of Galactic rotation ($\ell=90^{\circ},b=0^{\circ}$), and $W$ is positive toward the North Galactic Pole ($\ell=0^{\circ},b=90^{\circ}$).

Since the astrometric and spectroscopic observables are measured in the heliocentric reference frame, the derived Galactic space-velocity components are affected by the local effects of Galactic differential rotation. To account for this effect, corrections ($dU$ and $dV$) were applied to the initially calculated velocity components following the formalism of \citet{Mihalas1981} using the following relations:
\begin{eqnarray}
dU = V_{\rm rad} \cos \ell - v_{\rm t} \sin \ell,\\ \nonumber
dV = V_{\rm rad} \sin \ell - v_{\rm t} \cos \ell,\\ \nonumber
\end{eqnarray}
where $V_{\rm rad}$ and $v_{\rm t}$ denote the Galactic radial and tangential velocity components, respectively, and $\ell$ represents the Galactic longitude. First-order effects of Galactic rotation are confined to the in-plane velocity components ($U$ and $V$), while the vertical component $W$ remains unaffected, since no significant coupling exists between vertical motions and planar differential rotation in the solar neighborhood \citep{Coskunoglu2012, Bilir2012}. The corrections accounting for Galactic differential rotation are applied to the observed velocity components in the Galactic plane and are evaluated within the standard linear approximation. The corresponding kinematic corrections are expressed in terms of the radial and tangential components of the Galactic rotation field in the plane, commonly formulated via the Oort constants as
\begin{eqnarray}
v_{\rm t} &= r\times (A\times \cos 2\ell + B), \\ \nonumber
V_{\rm rad} &= A\times d\times \sin 2\ell. ~~~~~~~\\ \nonumber
\end{eqnarray}
where $A$ and $B$ are the Oort constants, for which we adopt $A=15.3\pm0.4$ and $B=-11.9\pm0.4\ {\rm km\,s^{-1}\,kpc^{-1}}$, as measured from Gaia DR1/TGAS data \citep{Bovy2017}. The parameter $r$ defines the projection distance of the star onto the Galactic plane, given by $r = d\times \cos b$, where $d$ is the heliocentric distance in kpc.

Finally, to eliminate heliocentric bias and project the motion onto the Local Standard of Rest (LSR) frame, we correct for the peculiar motion of the Sun. The absolute, corrected space velocity components relative to the LSR ($U_{\text{LSR}}, V_{\text{LSR}}, W_{\text{LSR}}$) are derived via linear translations:
\begin{eqnarray}
U_{\text{LSR}} &= U - \mathrm{dU} + U_{\odot}, \\ \nonumber
V_{\text{LSR}}, &= V - \mathrm{dV} + V_{\odot}, \\ \nonumber
W_{\text{LSR}} &= W + W_{\odot}, ~~~~~~\\ \nonumber
\end{eqnarray}
where we adopt the solar-motion vector $(U_{\odot}, V_{\odot}, W_{\odot}) = (8.83\pm0.24, 14.19\pm0.34, 6.57\pm0.21)\,\mathrm{km\,s^{-1}}$ \citep{Coskunoglu2011}. 
For the subsequent orbit calculations with \texttt{galpy}, the Solar-motion vector was transformed to the native \texttt{galpy} convention, for which the input is specified as $[-U_{\odot},V_{\odot},W_{\odot}]$. Accordingly, we adopted $(-8.83,14.19,6.57)~\mathrm{km\,s^{-1}}$ when initializing the orbits, ensuring consistency between the heliocentric velocity convention adopted above and the Galactocentric orbit calculations. The total space velocity relative to the LSR is then calculated as $S_{\rm LSR} = \sqrt{U_{\rm LSR}^2 + V_{\rm LSR}^2 + W_{\rm LSR}^2}$. To ensure a robust kinematic analysis, objects with uncertainties in $S_{\rm LSR}$ exceeding 20~km~s$^{-1}$ were excluded from the sample. This quality criterion removed 57 WDs, yielding a final kinematic sample of 5,837 objects used throughout the remainder of this study.

We further derive the Galactocentric velocity components in a cylindrical coordinate system with $V_{\rm R}$, $V_\phi$, and $V_{\rm Z}$, where $V_{\rm Z}$ is equivalent to $W$, and
\begin{eqnarray}
V_{\rm R} &=& \frac{1}{R} \left[ X_{\rm gc} \cdot U_{\rm LSR} + Y_{\rm gc} \cdot (V_{\rm LSR}+V_{\rm rot}) \right], \label{eqn:vrho}\\
V_\phi &=& \frac{1}{R} \left[ X_{\rm gc} \cdot (V_{\rm LSR}+V_{\rm rot}) - Y_{\rm gc} \cdot U_{\rm LSR} \right], \label{eqn:vphi}
\end{eqnarray}
and $R_{\rm gc}=\sqrt{X_{\rm gc}^2+Y_{\rm gc}^2}$ is the present-day Galactocentric cylindrical radius of each WD. We adopted a Solar Galactocentric distance of $R_0=8$~kpc and a local circular velocity of $V_{\rm rot}=220$~km~s$^{-1}$ \citep{Bovy2012, Bovy2015}. We calculate the position of each object in Galactocentric Cartesian coordinates ($X_{\rm gc}, Y_{\rm gc}, Z_{\rm gc}$)\footnote{The heliocentric space velocities are expressed in the right-handed convention defined above, whereas the orbital calculations follow the native left-handed Galactocentric convention of \texttt{galpy}.}:

\begin{eqnarray}
X_{\rm gc} = d \cos b\cos l - X_\odot, \\ \nonumber
Y_{\rm gc} = d \cos b\sin l, \\ \nonumber
Z_{\rm gc} = d \sin b. \\ \nonumber 
\end{eqnarray}

\subsection{Galactic Orbital Parameters}
\label{subsec:Galactic_orbit}

We used \texttt{galpy} \citep{Bovy2015} to derive the Galactic orbital parameters, adopting its standard axisymmetric \texttt{MWPotential2014} model. The potential consists of an exponentially truncated power-law bulge, a Miyamoto--Nagai disk \citep{Miyamoto1975}, and an NFW dark-matter halo \citep{Navarro1996}. We adopted the standard \texttt{MWPotential2014} scaling of $R_0=8$~kpc and $V_0=220$~km~s$^{-1}$. The bulge has a power-law index of 1.8 and a cutoff radius of 1.9~kpc; the disk has radial and vertical scale parameters of $a=3.0$~kpc and $b=0.28$~kpc, respectively; and the NFW halo has a scale radius of $r_{\rm s}=16$~kpc. The bulge, disk, and halo components are normalized to contribute 0.05, 0.60, and 0.35, respectively, to the radial force at $R_0$ \citep{Bovy2015}. We further assumed that the Sun lies $27\pm4$~pc above the Galactic plane \citep{Chen2001}.

Each orbit was integrated backward for 10~Gyr with a time step of 1~Myr. From the resulting trajectories, we derived the perigalactic ($R_{\rm peri}$) and apogalactic ($R_{\rm apo}$) distances, the maximum vertical distance from the Galactic plane ($Z_{\rm max}$), the orbital eccentricity ($e$), the guiding radius ($R_{\rm g}$), and the orbital period 
($P_{\rm orb}$).

\subsection{Kinematic Population Analysis}

We determine the Galactic population membership of each WD based on the kinematic classification scheme of \citet{Bensby2003, Bensby2005}, which assumes that the Galactic space-velocity components of the thin-disk ($D$), thick-disk ($TD$), and halo ($H$) populations follow Gaussian distributions characterized by distinct velocity dispersions and asymmetric drift velocities. For a given population $i$ ($D$, $TD$, and $H$) the probability distribution ($P_i$) is

\begin{equation}
\resizebox{\columnwidth}{!}{$
\label{equ:probability}
P_i(U,V,W) = k_i \exp \left[
-\frac{U_{\rm LSR}^2}{2\sigma_{U,i}^2}
-\frac{(V_{\rm LSR}-V_{{\rm asym},i})^2}{2\sigma_{V,i}^2}
-\frac{W_{\rm LSR}^2}{2\sigma_{W,i}^2}
\right],
$}
\end{equation}
where $\sigma_{\rm U}$, $\sigma_{\rm V}$, and $\sigma_{\rm W}$ are the characteristic velocity dispersions, $V_{\rm asym}$ is the asymmetric drift velocity, and $k_i$ is the normalization constant,

\begin{equation}
k_i = \frac{1}{(2\pi)^{3/2}\sigma_{\rm U}\sigma_{\rm V}\sigma_{\rm W}}.
\end{equation}
Following \citet{Bensby2005}, the characteristic velocity dispersions $(\sigma_{\rm U},\sigma_{\rm V},\sigma_{\rm W})$ are $(35,20,16)$~km~s$^{-1}$ for the thin disk, $(67,38,35)$~km~s$^{-1}$ for the thick disk, and $(160,90,90)$~km~s$^{-1}$ for the halo. The corresponding asymmetric drift velocities, $V_{\rm asym}$, are $-15$, $-46$, and $-220$~km~s$^{-1}$, respectively. Local stellar population fractions of $X_{\rm D}=0.9385$, $X_{\rm TD}=0.06$, and $X_{\rm H}=0.0015$ were adopted based on the estimates from \citet{Buser1999} and \citet{Bilir2008}. The relative probabilities between populations were computed as

\begin{equation}
\frac{TD}{D} =
\frac{X_{\rm TD}}{X_{\rm D}}
\frac{P_{\rm TD}}{P_{\rm D}},
\qquad
\frac{TD}{H} =
\frac{X_{\rm TD}}{X_{\rm H}}
\frac{P_{\rm TD}}{P_{\rm H}},
\end{equation}
where $P_i$ denotes the probability associated with population $i$. We classify stars with $TD/D \leq 0.1$ as high-probability thin-disk members, and those with $TD/D > 10$ as high-probability thick-disk or halo stars. We identify objects with $0.1 < TD/D \leq 1$ and $1 < TD/D \leq 10$ as low-probability thin-disk and thick-disk members, respectively. For the population-based analyses presented in this study, the high- and low-probability thin-disk categories were combined, and WDs with $TD/D\leq1$ were assigned to the thin-disk sample. Among the objects with $TD/D>10$, those satisfying $TD/H>1$ were classified as thick-disk members, whereas those with $TD/H\leq1$ were assigned to the halo. Thus, the $TD/H$ ratio explicitly distinguishes the thick-disk and halo populations within the kinematically hot ($TD/D>10$) subsample. Objects with $1<TD/D\leq10$ were classified as transition objects and were therefore excluded from the final thin-disk, thick-disk, and halo samples.


\subsection{Jeans Analysis of Radial and Vertical Disc Scales from White Dwarfs}
\label{sec:Jeans}

Although WDs are intrinsically faint and therefore restricted to a relatively small volume in the Solar neighborhood, their kinematics still encode information on the global dynamical properties of the Galactic disk. Thus, the observed velocity ellipsoids can be exploited within the Jeans formalism to obtain indirect constraints on the structural parameters of the thin and thick disk components \citep{Kordopatis2011}. The geometric distances adopted from \citet{Bailer-Jones2021} are based on a direction-dependent Galactic distance prior. This prior does not directly set the disk scale lengths or scale heights derived here, because these quantities are inferred from the Jeans equations using the velocity moments and their spatial gradients rather than from a direct fit to the stellar distance distribution. Nevertheless, the adopted distances enter indirectly through the Galactocentric positions and tangential velocities, and therefore some residual dependence on the distance prior cannot be excluded. The effect is expected to be limited for our nearby sample, which has a median distance of 214 pc and is restricted to sources with relative distance uncertainties below 20\%. In the following, we combine the derived velocity ellipsoids with the Jeans equations under the assumption of steady-state equilibrium to estimate 
the radial scale length ($h_{\rm R}$) and the vertical scale height ($h_{\rm Z}$). In cylindrical coordinates, the Jeans equations for the radial and vertical components are given by
\begin{equation}
v_{\rm c}^2-\overline{v_\phi}^2=\sigma^2_{V_{\rm R}} \left(\frac{\sigma^2_{V_\phi}}{\sigma^2_{V_{\rm R}}} -1 - \frac{\partial \ln(\rho \sigma^2_{V_{\rm R}})}{\partial \ln R} - \frac{R}{\sigma^2_{V_{\rm R}}} \frac{\partial \sigma^2_{V_{\rm RZ}}}{\partial Z} \right),
\label{equ: scale-lenght}
\end{equation}

\begin{equation}
\rho K_{\rm Z} = \frac{\partial (\rho \sigma^2_{V_{\rm Z}})}{\partial Z} + \frac{1}{R} \frac{\partial (R \rho \sigma^2_{V_{\rm RZ}})}{\partial R}.
\label{eqn:vertical_jeans_general}
\end{equation}
Here, $\rho$ denotes the spatial density of the Galactic component. The circular velocity at the Solar radius is adopted as $V_{\rm c}=220~\mbox{km s}^{-1}$, and $\overline{V_\phi}$ is the mean azimuthal velocity of the stellar population characterized by the velocity dispersions $\sigma_{V_{\rm R}}$, $\sigma_{V_\phi}$, and $\sigma_{V_{\rm Z}}$. The cross-term of the velocity dispersion tensor is defined as $\sigma^2_{V_{\rm RZ}}=\overline{V_{\rm R}V_{\rm Z}}-\overline{V_{\rm R}}~\overline{V_{\rm Z}}$, while $K_{\rm Z}$ represents the vertical component of the Galactic gravitational acceleration.

\subsubsection{Radial scale lengths}
We adopt an exponential radial density profile of the form $\rho(R) \propto \exp(-R/h_{\rm R})$, and assume that the radial velocity dispersion follows the same functional dependence as the density distribution \citep[e.g.][]{Binney2008, Bovy2013}. With this assumption, the radial velocity dispersion can be written as $\sigma^2_{V_{\rm R}} \propto \exp(-R/h_{\rm R})$. Furthermore, we neglect the vertical gradient of the cross term, i.e., $\partial \sigma^2_{V_{\rm R, Z}}/\partial Z \approx 0$, which is valid for a locally plane-parallel potential corresponding to an infinite constant surface density sheet \citep{Gilmore1989}. In this limit, the principal axes of the velocity ellipsoid are aligned with the cylindrical coordinate system, and Equation~(\ref{equ: scale-lenght}) reduces to
\begin{equation}
\frac{\sigma^2_{V_\phi}}{\sigma^2_{V_R}} -1 + \frac{2R}{h_R} - \frac{v_c^2-\overline{v_\phi}^2}{\sigma^2_{V_R}}=0 .
\label{eqn:jeans1}
\end{equation}

Alternatively, if the Galactic potential is dominated by a centrally concentrated mass distribution, the local velocity ellipsoid is expected to be oriented towards the Galactic center \citep{Gilmore1989, Binney2008}, where the cross-term contribution becomes
\begin{equation}
\frac{r}{\sigma^2_{V_R}} \frac{\partial \sigma^2_{V_{R,Z}}}{\partial Z} \approx 1 - \frac{\sigma^2_{V_Z}}{\sigma^2_{V_R}} .
\label{eqn:approx_Gilmore}
\end{equation}
Substituting this approximation into Equation~(\ref{equ: scale-lenght}) yields
\begin{equation}
\frac{\sigma^2_{V_\phi}}{\sigma^2_{V_{\rm R}}} -2 + \frac{2R}{h_{\rm R}} - \frac{v_{\rm c}^2-\overline{v_\phi}^2}{\sigma^2_{V_{\rm R}}} + \frac{\sigma^2_{V_{\rm Z}}}{\sigma^2_{V_{\rm R}}}=0.
\label{eqn:jeans2}
\end{equation}
All terms appearing in Equations~(\ref{eqn:jeans1}) and (\ref{eqn:jeans2}) are directly constrained by the observational data, leaving the radial scale length $h_{\rm R}$ as the only free parameter of the disk model.

\subsubsection{Vertical scale heights}

Following \citet{Kordopatis2011}, we neglect the final term in Equation~(\ref{eqn:vertical_jeans_general}), 
$\frac{1}{R}\frac{\partial(R\rho\sigma^2_{V_{\rm RZ}})}{\partial R}$, which represents the radial variation of the cross term of the velocity 
dispersion tensor. This approximation is adopted because our sample lies in the Solar neighborhood, far from the Galactic center, where this contribution 
is expected to be small. We further assume that the vertical density profile of the disc component follows an exponential form, $\rho(Z) \propto \exp(-Z/h_{\rm Z})$. Under these assumptions, Equation~(\ref{eqn:vertical_jeans_general}) can be rearranged into a more compact form as
\begin{equation}
\frac{\partial \ln \sigma^2_{V_Z}}{\partial Z} + \frac{K_{\rm Z}}{\sigma^2_{V_{\rm Z}}} - \frac{1}{h_{\rm Z}} = 0,
\label{eqn:vertical_jeans}
\end{equation}
which directly relates the vertical gradient of the velocity dispersion to the local gravitational force and the vertical scale height of the Galactic disk.

\section{Results}
\label{sec:result}

\subsection{Spatial Distributions}

\begin{figure}[t]
\centering
\includegraphics[width=0.98\columnwidth]{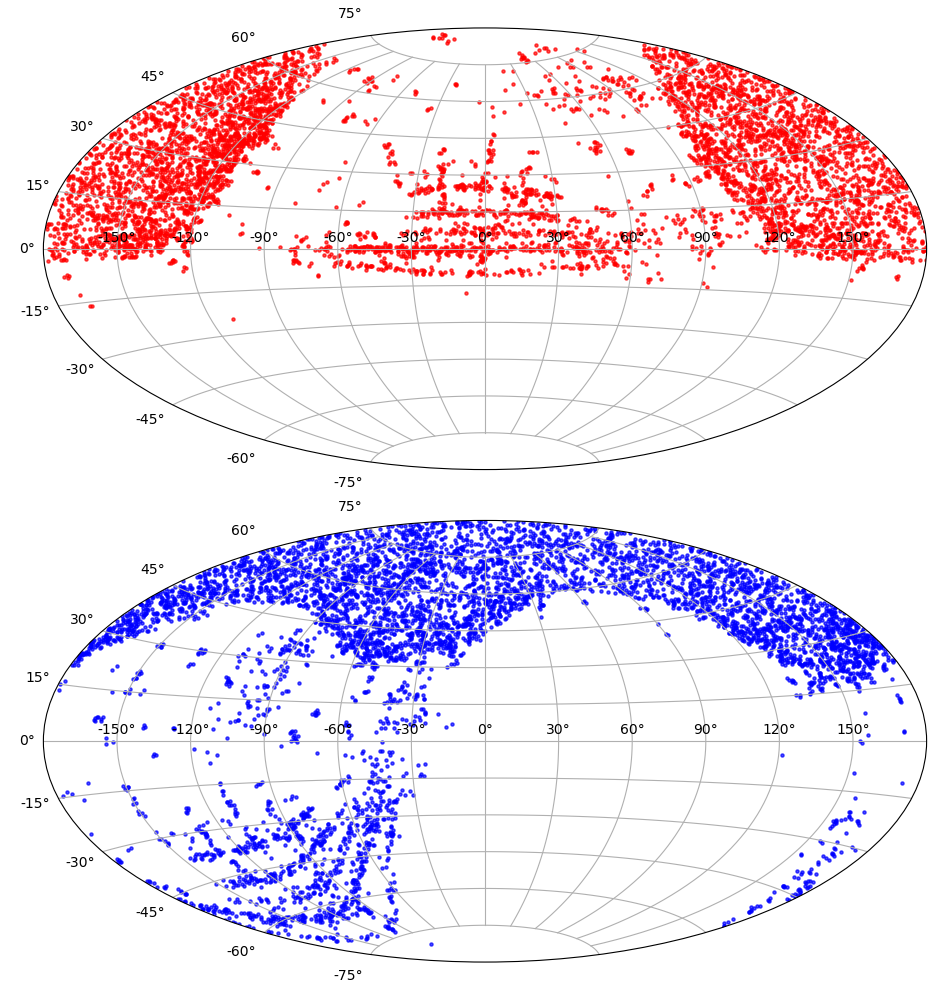}
\caption{Aitoff projections of our WD sample in the equatorial ($\alpha,\delta$; top panel) and Galactic ($l,b$; bottom panel) coordinate systems.} 
\label{fig:aitoff}
\end {figure}

Figure~\ref{fig:aitoff} presents the Aitoff projections of the positions of our sample of white dwarfs in both equatorial and Galactic coordinate systems. The equatorial distribution reflects the observational footprint of the SDSS, which was conducted primarily from the Apache Point Observatory in the northern hemisphere. Not surprisingly, $\approx91\%$ of the WDs in our sample are located in the northern celestial hemisphere. The distribution becomes slightly more balanced in Galactic coordinates, with approximately $81\%$ of the objects located at positive Galactic latitudes. Although the sample is subject to the selection effects imposed by the SDSS survey geometry, WDs are represented on both sides of the Galactic plane. Furthermore, the SDSS observing strategy preferentially targeted regions with Galactic latitudes of $|b|>20^{\circ}$.

\begin{figure}[t]
\centering
\includegraphics[width=0.98\columnwidth]{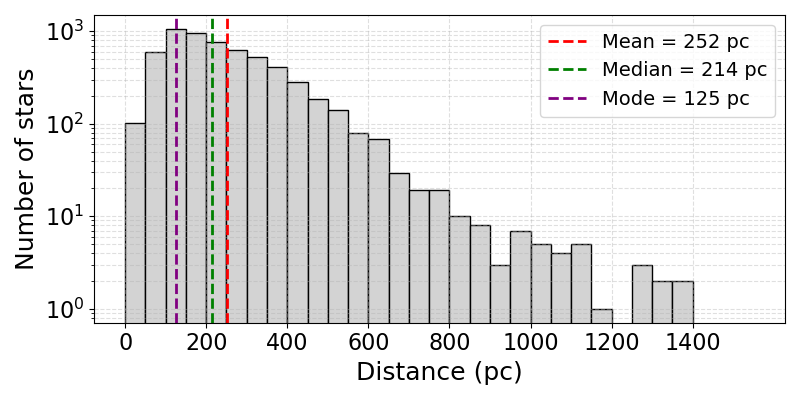}
\caption{Histogram of the heliocentric distances of the WDs in our sample.} 
\label{fig:distance_hist}
\end {figure}

Figure~\ref{fig:distance_hist} shows the distance distribution of our sample based on \citet{Bailer-Jones2021} estimates. Since SDSS spectroscopy is a magnitude-limited survey, we find a skewed distribution characterized by a concentration of nearby objects and a progressively decreasing number of sources at larger distances. The mean, median, and mode of the distribution are 252, 214, and 125 pc, respectively. 

\begin{figure}[t]
\centering
\includegraphics[width=0.98\columnwidth]{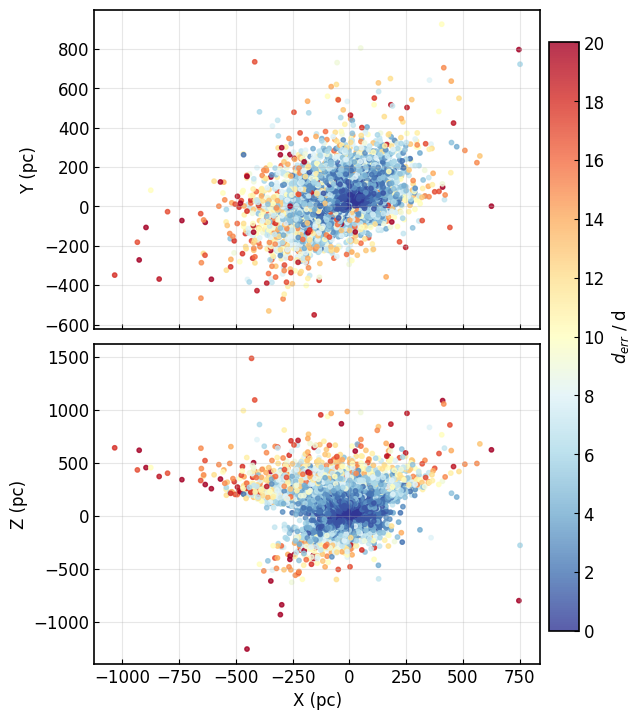}
\caption{Distribution of the WDs in the heliocentric Galactic Cartesian $X-Y$ (top) and $X-Z$ (bottom) planes. The color scale indicates the relative distance uncertainty.} 
\label{fig:xyz}
\end {figure}

Figure~\ref{fig:xyz} shows the distribution of our sample in the heliocentric Galactic Cartesian coordinates ($X$, $Y$, $Z$). Symbols are color-coded according to their relative distance uncertainties ($d_{\rm err}/d$). Our WD sample is roughly symmetrically distributed around the Sun in the galactocentric $X$ and $Y$ coordinates, where the median positions of the sample are found to be at $X=-35$~pc, $Y=26$~pc, and $Z=125$~pc. The positive median $Z$ value indicates that the sample is distributed preferentially above the Galactic plane. This asymmetry is not expected to reflect the intrinsic distribution of nearby WDs, but is instead primarily a consequence of the SDSS survey geometry. For comparison, \citet{Raddi2022} also found a similar offset in $Z$ of 120 pc using the SDSS DR12 DA WD sample. 

\subsection{Kinematics}
\label{sec:kinematics}

\begin{figure*}[ht]
\centering
\includegraphics[width=0.95\textwidth]{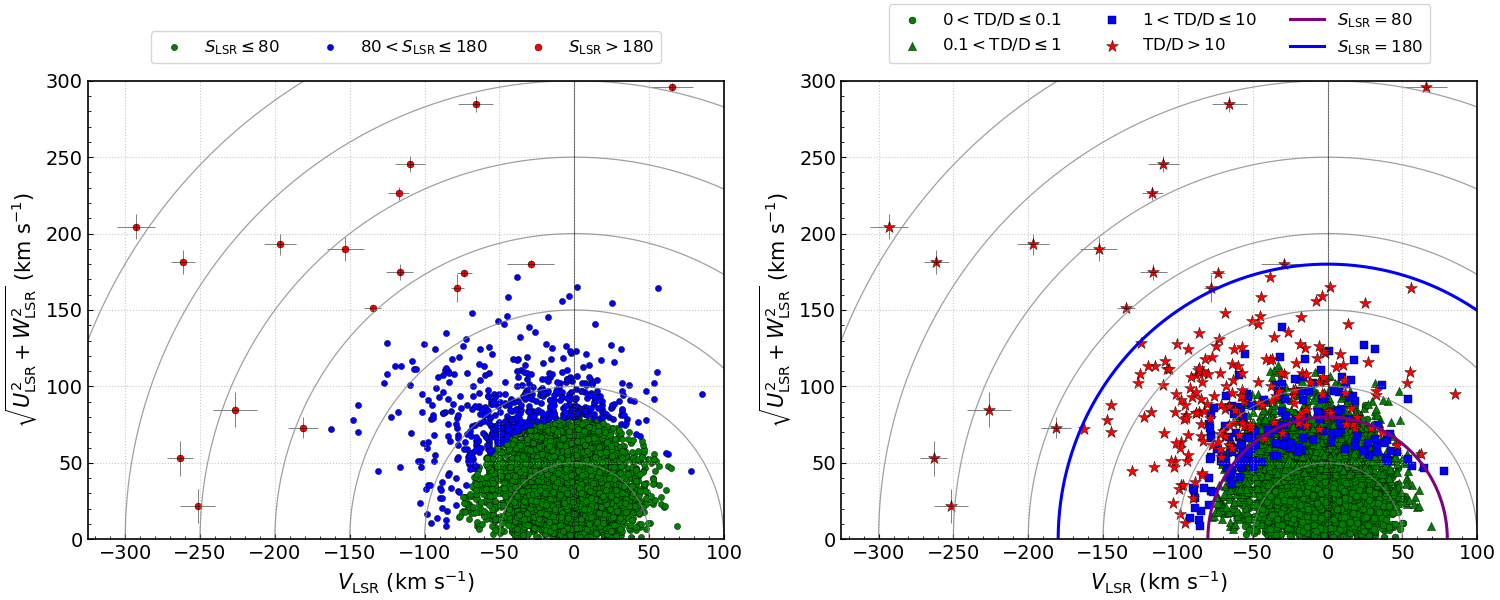}
\caption{Toomre diagrams for our WD sample. The two panels show the Galactic population classification based on $S_{\rm tot}$ \citep[][left]{Nissen2004} and the kinematic criteria of \citet[][right]{Bensby2003, Bensby2005}. The gray curves represent iso-total-velocity contours, while the blue and purple curves denote the velocity boundaries adopted to distinguish the thick- and thin-disk populations, respectively. } 
\label{fig:toomre}
\end {figure*}

Figure~\ref{fig:toomre} presents the Toomre diagrams constructed for the final sample of 5,837 WDs after applying the gravitational redshift, Galactic differential rotation, and LSR corrections. To assess the Galactic population membership of the sample, we consider two widely used kinematic classification schemes. The first adopts fixed total space velocity thresholds following \citet{Nissen2004}, whereas the second employs the probability-based formalism of \citet{Bensby2003, Bensby2005}. The two approaches provide complementary views of the kinematic structure of the sample and enable a direct comparison between deterministic and probabilistic population assignments. Using the velocity limits proposed by \citet{Nissen2004}, we classify stars with $S_{\rm LSR}\le80$ km s$^{-1}$ as thin-disk members, objects with $80<S_{\rm LSR}~{\rm (km~s^{-1})}\le180$ as thick-disk stars, and those with $S_{\rm LSR}>180$ km s$^{-1}$ as halo candidates. According to these criteria, 5,304 ($90.9\%$) WDs belong to the thin disk, 516 ($8.8\%$) to the thick disk, and only 17 ($0.3\%$) satisfy the halo criterion. The resulting Toomre diagram exhibits the expected kinematic sequence, with thin-disk stars concentrated near the origin and progressively higher velocity populations occupying the outer regions of the diagram.

The probabilistic classification method of \citet{Bensby2003, Bensby2005} accounts for the overlap between the velocity distributions of different Galactic populations. Based on the $TD/D$ probability ratio, we classify 4,840 WDs as high-probability thin-disk members ($TD/D\leq0.1$), 628 as low-probability thin-disk members ($0.1<TD/D\leq1$), and 167 as transition objects between the thin and thick disks ($1<TD/D\leq10$). For the 202 WDs with $TD/D>10$, we use the $TD/H$ ratio to distinguish thick-disk from halo members. Of these, 186 WDs with $TD/H>1$ are assigned to the thick disk, while 16 with $TD/H\leq1$ are classified as halo members. We adopt the classification scheme of \citet{Bensby2003, Bensby2005} throughout the remainder of this study.  The high- and low-probability thin-disk categories are combined, yielding a final thin-disk sample of 5,468 WDs with $TD/D\leq1$. The 167 objects with $1<TD/D\leq10$ are treated as kinematic transition objects and are not included in any of the three final population samples. Our classified sample therefore consists of 5,468 thin-disk, 186 thick-disk, and 16 halo WDs. Among the 5,670 WDs assigned to a specific Galactic population, these numbers correspond to observed fractions of $96.4\%$, $3.3\%$, and $0.3\%$, respectively. Thus, the solar-neighborhood WD sample examined here is overwhelmingly dominated by the thin-disk population. A similarly small halo contribution ($\sim1$\%) was recently reported by \citet{Swan2026} for the DESI DR1 WD sample. Although their probabilistic disk--halo decomposition differs from the population criteria adopted here, both analyses consistently indicate that disk stars overwhelmingly dominate the local WD population, with only a minor contribution from the halo.

\subsection{Galactic Orbital Properties}

The orbital integrations described in Section~\ref{subsec:Galactic_orbit} provide the Galactic orbital parameters ($R_{\rm peri}$, $R_{\rm apo}$, $Z_{\rm max}$, $e$, $R_{\rm g}$, and $P_{\rm orb}$) for all 5,837 WDs in the final kinematic sample. The equatorial and Galactic coordinates, velocity components, orbital parameters, total ages, and progenitor masses for a representative subset of the studied WDs are listed in Appendix~\ref{sec:appendix} (Table~\ref{tab:Data}), while the complete catalog is available in the online supplementary material.

The orbital parameter distributions reveal that the sample spans a broad range of dynamical properties. Most WDs follow nearly circular orbits confined to the Galactic disk, as indicated by their low eccentricities and small values of $Z_{\rm max}$. Nevertheless, a small fraction of the sample reaches several kpc above the Galactic plane, with a few objects attaining $Z_{\rm max}\sim10$~kpc. While the eccentricity distribution is strongly peaked toward low values, a limited number of stars exhibit highly eccentric orbits. These objects are consistent with dynamically hotter populations associated with the thick disk and halo.

\begin{figure}[b]
\centering
\includegraphics[width=1\columnwidth]{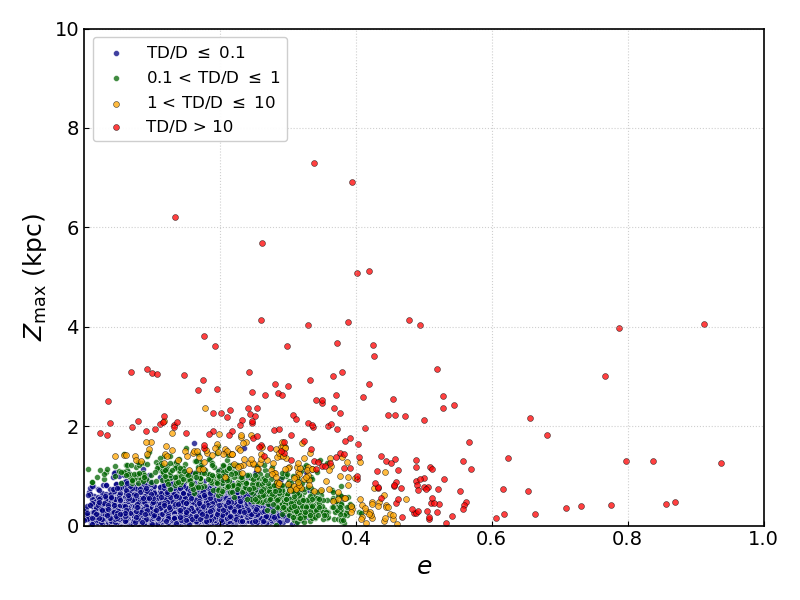}
\caption{$Z_{\rm max}$ vs eccentricity for our WD sample. Each object is color-coded according to its population membership based on the kinematic criteria of \citet{Bensby2003, Bensby2005}.} 
\label{fig:ze}
\end {figure}

\begin{table*}[t]
\setlength{\tabcolsep}{2.8pt}
\renewcommand{\arraystretch}{1.2}
\centering
\tiny
\caption{Mean space velocity components and their dispersions, total ages, mean progenitor masses, and mean Galactic orbital eccentricities of the WDs classified into Galactic populations according to four different criteria.}
\label{tab:Kinematics}
\begin{tabular}{lrrrrrrrrrcc}
\hline
\hline
Criterion & \multicolumn{1}{c}{$N$} & \multicolumn{1}{c}{$U_{\rm LSR}$} & \multicolumn{1}{c}{$V_{\rm LSR}$} & \multicolumn{1}{c}{$W_{\rm LSR}$} & \multicolumn{1}{c}{$\sigma_{\rm U}$} & \multicolumn{1}{c}{$\sigma_{\rm V}$} & \multicolumn{1}{c}{$\sigma_{\rm W}$} & \multicolumn{1}{c}{$\sigma_{\rm tot}$} & $\langle t_{\rm tot}\rangle$ & $\langle M\rangle$ & $\langle e\rangle$ \\ 
\cline{3-9}
 &  & \multicolumn{7}{c}{(km s$^{-1}$)} & (Gyr) & ($M_{\rm \odot}$) &  \\ 
\hline
All sample & 5,837 & $-0.44 \pm 4.30$ & $-7.48 \pm 3.76$ & $-1.47 \pm 6.26$ & $38.10 \pm 0.35$ & $27.66 \pm 0.26$ & $24.37 \pm 0.23$ & $53.01 \pm 0.49$ & 4.42 & 1.66 & 0.16 \\
\hline
$S_{\rm tot} \le 80$ & 5,304 & $0.14 \pm 4.18$ & $-4.21 \pm 3.65$ & $-1.67 \pm 6.18$ & $31.39 \pm 0.30$ & $22.31 \pm 0.22$ & $19.88 \pm 0.19$ & $43.34 \pm 0.42$ &  4.36 & 1.67 & 0.14 \\
$80 < S_{\rm tot} \le 180$ & 516 & $-5.46 \pm 5.49$ & $-36.50 \pm 4.76$ & $2.01 \pm 7.07$ & $75.25 \pm 2.34$ & $40.81 \pm 1.27$ & $45.96 \pm 1.43$ & $97.16 \pm 3.03$ & 5.10 & 1.56 & 0.34 \\
$S_{\rm tot} > 180$ & 17 & $-29.74 \pm 5.69$ & $-146.19 \pm 6.51$ & $-44.41 \pm 7.68$ & $131.84 \pm 23.31$ & $93.29 \pm 16.49$ & $119.62 \pm 21.15$ & $200.98 \pm 35.53$ & 5.01 & 1.45 & 0.66 \\
\hline
$TD/D \le 0.1$ & 4,840 & $0.82 \pm 4.19$ & $-3.35 \pm 3.59$ & $-1.25 \pm 6.09$ & $32.33 \pm 0.33$ & $19.64 \pm 0.20$ & $16.03 \pm 0.16$ & $41.08 \pm 0.42$ & 4.33 & 1.67 & 0.13 \\
$0.1<TD/D\le1$ &   628 & $-4.27 \pm 4.75$ & $-16.50 \pm 4.34$ & $-4.78 \pm 6.89$ & $51.37 \pm 1.45$ & $36.59 \pm 1.03$ & $33.59 \pm 0.95$ & $71.46 \pm 2.02$ & 4.93 & 1.54 & 0.25 \\
$1<TD/D<10$    &   167 & $-10.93 \pm 4.72$ & $-27.94 \pm 4.60$ & $1.27 \pm 7.13$ & $56.49 \pm 3.10$ & $40.23 \pm 2.21$ & $46.31 \pm 2.54$ & $83.39 \pm 4.58$ & 4.92 & 1.63 & 0.29 \\
$TD/D>10$      &   202 & $-9.95 \pm 5.20$ & $-61.47 \pm 5.31$ & $1.48 \pm 7.67$ & $76.17 \pm 3.80$ & $58.65 \pm 2.93$ & $75.31 \pm 3.76$ & $122.12 \pm 6.09$ & 4.78 & 1.67 & 0.39 \\
Thin disk      & 5,468 & $0.23 \pm 4.26$ & $-4.86 \pm 3.68$ & $-1.66 \pm 6.18$ & $35.08 \pm 0.34$ & $22.64 \pm 0.22$ & $18.93 \pm 0.18$ & $45.84 \pm 0.44$ & 4.40 & 1.66 & 0.15 \\
Thick disk     &   186 & $-10.71 \pm 5.20$ & $-54.33 \pm 5.16$ & $3.56 \pm 7.59$ & $70.72 \pm 3.68$ & $46.63 \pm 2.42$ & $67.02 \pm 3.48$ & $108.02 \pm 5.62$ & 4.71 & 1.69 & 0.38 \\
Halo           &   16  & $-1.08 \pm 5.24$ & $-144.52 \pm 7.08$ & $22.80 \pm 8.63$ & $122.53 \pm 22.37$ & $103.29 \pm 18.86$ & $136.90 \pm 24.99$ & $210.77 \pm 38.48$ & 4.41 & 1.40 & 0.65 \\
\hline
$Z_{\max}\le1$ & 5,408 & $-0.19 \pm 4.26$ & $-6.74 \pm 3.72$ & $-1.61 \pm 6.14$ & $36.26 \pm 0.35$ & $25.31 \pm 0.24$ & $17.72 \pm 0.17$ & $47.64 \pm 0.46$ & 4.38 & 1.66 & 0.15 \\
$1<Z_{\max}\le4$ & 412 & $-3.78 \pm 4.88$ & $-15.37 \pm 4.29$ & $1.16 \pm 7.77$ & $54.36 \pm 1.90$ & $43.03 \pm 1.50$ & $57.67 \pm 2.01$ & $90.18 \pm 3.15$ & 5.05 & 1.59 & 0.23 \\
$Z_{\max}>4$ & 17 & $1.94 \pm 3.62$ & $-53.49 \pm 5.07$ & $-19.05 \pm 8.43$ & $90.97 \pm 16.08$ & $100.41 \pm 17.75$ & $151.39 \pm 26.76$ & $203.17 \pm 35.92$ & 3.72 & 2.17 & 0.44 \\
\hline
\hline
\end{tabular}
\end{table*}

Figure~\ref{fig:ze} shows the distribution of the sample in the $Z_{\rm max}$ vs $e$ plane, where the kinematic populations identified using the \citet{Bensby2003, Bensby2005} criteria are indicated. As expected, thin-disk WDs are concentrated at low eccentricities and short vertical distances, whereas thick-disk and halo candidates occupy progressively larger regions of the diagram. However, a small number of stars exhibit relatively large eccentricities despite remaining confined to $Z_{\rm max}<1$~kpc. These objects possess dynamically hot radial orbits while remaining close to the Galactic plane, suggesting that radial migration and orbital blurring have played an important role in shaping their present-day orbital properties. 

\subsection{Velocity Dispersions and Population Properties}

Table~\ref{tab:Kinematics} presents the Galactic velocity components and their dispersions, total stellar ages, progenitor masses, and orbital eccentricities for the different subsamples. For the entire sample, the mean velocity components relative to the LSR are $\langle U_{\rm LSR}\rangle=-0.4\pm4.3$, $\langle V_{\rm LSR}\rangle=-7.5\pm3.8$, and $\langle W_{\rm LSR}\rangle=-1.5\pm6.3$ km~s$^{-1}$. The corresponding velocity dispersions are $\sigma_{\rm U}=38.1\pm0.4$, $\sigma_{\rm V}=27.7\pm0.3$, and $\sigma_{\rm W}=24.4\pm0.2$ km~s$^{-1}$, yielding a total velocity dispersion of $\sigma_{\rm tot}=53.0\pm0.5$ km~s$^{-1}$. The measured velocity dispersions define an anisotropic velocity ellipsoid typical of the Galactic disk. Such a distribution is consistent with the secular dynamical evolution of the Galactic disk, in which stellar random motions are amplified more efficiently in the radial direction than perpendicular to the disk. For the complete sample, the median total age, progenitor mass, and Galactic orbital eccentricity are 4.4 Gyr, 1.66 $M_\odot$, and 0.16, respectively. The median age and low orbital eccentricity are consistent with the predominance of thin-disk WDs in the sample, while the inferred progenitor mass corresponds to an intermediate-mass main-sequence star.

When the sample is divided according to the total space velocity \citep{Nissen2004}, the total velocity dispersion increases steadily from $\sigma_{\rm tot} = 43.3 \pm 0.4$ km~s$^{-1}$ for stars with $S_{\rm LSR}\leq80$ km~s$^{-1}$ to $201.0 \pm 35.5$ km~s$^{-1}$ for the highest-velocity subsample ($S_{\rm LSR}>180$ km~s$^{-1}$). This increase is accompanied by progressively larger dispersions in all three velocity components and by a pronounced decrease in the mean rotational velocity, reaching $\langle V_{\rm LSR}\rangle=-146.2 \pm 6.5$ km~s$^{-1}$ for the fastest stars, indicative of a substantial rotational lag with respect to the Galactic disk. At the same time, the median total age of the WDs increases from 4.4 to 5.0 Gyr, while the median progenitor mass decreases from 1.67 to 1.45 $M_{\odot}$, and the median Galactic orbital eccentricity rises from 0.14 to 0.66. These systematic trends indicate a gradual transition from a sample dominated by dynamically cold thin-disk WDs to a population with larger contributions from the thick disk and the halo. The increasing Galactic orbital eccentricities reflect the transition from nearly circular disk orbits to the more elliptical orbits characteristic of dynamically hotter Galactic populations.

A similar kinematic progression is recovered when the WDs are classified as thin-disk, thick-disk, and halo members using the population probabilities of \citet{Bensby2003, Bensby2005}. Objects with $1<TD/D\leq 10$ were classified as transition objects and excluded from the final population samples. The thin-disk population, comprising 5,468 WDs, has mean space velocities of $(U_{\rm LSR},V_{\rm LSR},W_{\rm LSR})=(0.2,-4.9,-1.7)$ km~s$^{-1}$ and exhibits the smallest velocity dispersions, with $(\sigma_{\rm U},\sigma_{\rm V},\sigma_{\rm W})=(35.1, 22.6, 18.9)$ km~s$^{-1}$ and $\sigma_{\rm tot}=45.8\pm0.4$ km~s$^{-1}$. For the 186 thick-disk WDs, the corresponding mean space velocity components are $(U_{\rm LSR},V_{\rm LSR},W_{\rm LSR})=(-10.7,-54.3,3.6)$ km~s$^{-1}$, while the space velocity component dispersions increase to $(\sigma_{\rm U},\sigma_{\rm V},\sigma_{\rm W})=(70.7, 46.6, 67.0)$ km~s$^{-1}$, yielding $\sigma_{\rm tot}=108.0\pm5.6$ km~s$^{-1}$. The 16 halo WDs show the largest rotational lag, with $(U_{\rm LSR},V_{\rm LSR},W_{\rm LSR})=(-1.1, -144.5, 22.8)$ km~s$^{-1}$, and the broadest velocity distribution, characterized by $(\sigma_{\rm U},\sigma_{\rm V},\sigma_{\rm W})=(122.5,103.3,136.9)$ km~s$^{-1}$ and $\sigma_{\rm tot}=210.8\pm38.5$ km~s$^{-1}$. Thus, the increasingly negative mean $V_{\rm LSR}$ and the systematic growth of both the component and total velocity dispersions from the thin disk to the halo reflect progressively hotter stellar kinematics and a larger rotational lag. This trend is particularly pronounced in $\sigma_{\rm W}$, indicating progressively stronger vertical dynamical heating from the thin disk to the halo. The orbital eccentricities show a consistent progression, with their mean value increasing from 0.15 for the thin disk to 0.38 for the thick disk and 0.65 for the halo.

A consistent picture is also obtained when the WDs are classified according to their maximum distance from the Galactic plane ($Z_{\rm max}$). Following the adopted criteria, stars with $Z_{\rm max}\leq1$ kpc, $1< Z_{\rm max}~{\rm (kpc)}\leq 4$, and $Z_{\rm max}>4$ kpc are assigned to the thin disk, thick disk, and halo populations, respectively. The total velocity dispersion increases systematically from $\sigma_{\rm tot}=47.6$ km~s$^{-1}$ for the thin-disk population to $\sigma_{\rm tot}=203.2$ km~s$^{-1}$ for the halo population. In contrast, the median total ages do not exhibit a monotonic variation across the three Galactic populations, even though the Galactic orbital eccentricities increase steadily from 0.15 for the thin-disk population to 0.44 for the halo population. Overall, the kinematic trends obtained from the $Z_{\rm max}$-based subsamples are consistent with those inferred from the velocity- and probability-based classifications. Both the velocity dispersions and orbital eccentricities increase systematically from the thin disk to the halo, whereas neither the median total age nor the progenitor mass exhibits a similarly robust correlation with Galactic population. The absence of a clear age or progenitor-mass sequence should, however, be interpreted with caution, since the halo subsample contains only a limited number of objects, reducing the statistical significance of any evolutionary trends involving this population.

\subsection{Age-Velocity Dispersion Relations}

\begin{figure}[t]
\centering
\includegraphics[width=1\columnwidth]{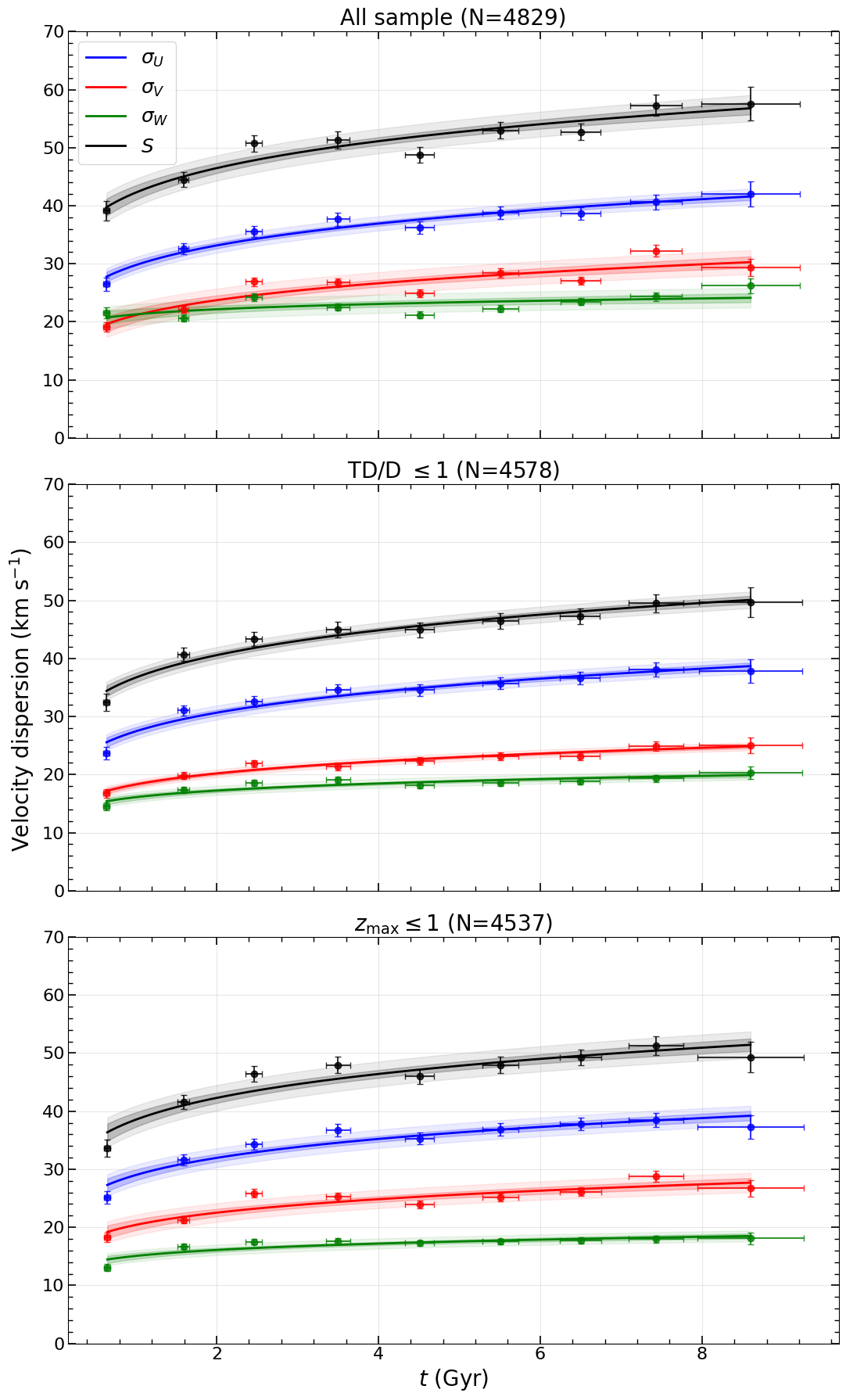}
\caption{Velocity dispersion vs age for the entire WD sample (top), those classified as thin-disk members ($TD/D\leq1$) according to the kinematic criteria of \citet[][middle]{Bensby2003, Bensby2005}, and those with $Z_{\rm max}\leq1$ kpc (bottom).} 
\label{fig:age-dis}
\end {figure}

To investigate the age dependence of WD kinematics, we analyzed the AVR using the subset of 4,829 WDs for which total stellar ages could be determined with the \texttt{wdwarfdate} package \citep{Kiman2022}. To assess the effect of population contamination, the analysis was carried out on three samples: (1) the complete sample with available total stellar ages ($N=4,829$), (2) a kinematically selected thin-disk sample ($N=4,578$) satisfying $TD/D \leq 1$ \citep{Bensby2003, Bensby2005}, and (3) a sample ($N=4,537$) restricted to $Z_{\rm max} \leq 1$~kpc. In each sample, stars were grouped into 1~Gyr age bins. For each age bin, the velocity dispersions $\sigma_{\rm U}$, $\sigma_{\rm  V}$, and $\sigma_{\rm  W}$ were calculated as the sample standard deviations of the corresponding $U_{\rm LSR}$, $V_{\rm LSR}$, and $W_{\rm LSR}$ velocity components, while $\sigma_{\rm tot}$ was calculated from $S_{\rm LSR}$. The median age of each bin was adopted as the representative age to minimize the influence of outliers. The uncertainty in each velocity dispersion was estimated from the standard error of the sample standard deviation, $\epsilon_{\sigma_i}=\sigma_i/\sqrt{2(N-1)}$, where $N$ is the number of WDs in the corresponding age bin \citep{Joncour2017}. The AVR relation $\sigma_i=A\times t^B$ was fitted to the binned velocity dispersions using an unweighted fit, and the goodness of each fit was quantified by the coefficient of determination ($R^2$). Individual WD total ages can have substantial and asymmetric uncertainties, particularly because uncertainties in the WD parameters propagate into the inferred progenitor masses and pre-WD evolutionary lifetimes. In some cases, the upper uncertainty limits may extend beyond the age of the Galaxy. We therefore interpret the AVR on a statistical, population-wide basis rather than assigning significance to the age of any individual WD. The use of 1~Gyr age bins and the median age within each bin reduces the sensitivity to individual extreme age estimates; nevertheless, age uncertainties may broaden the distributions and contribute to the observed scatter in the AVR.

\begin{table*}[htbp]
\centering
\caption{Total numbers of WDs, mean ages, and space velocity dispersions for different total stellar age bins.}
\label{tab:age-kinematic}
\begin{tabular}{cccccccc}
\hline\hline
Sample        & Age range & $N$ & $\langle t\rangle$ & $\sigma_{\rm U}$ & $\sigma_{\rm V}$ & $\sigma_{\rm W}$ & $\sigma_{\rm tot}$ \\
\cline{5-8}
              & (Gyr)     &     &  (Gyr)             &\multicolumn{4}{c}{(km s$^{-1}$)} \\
\hline
              & 0--1 & 271 & $0.64 \pm 0.04$ & $26.48 \pm 1.14$ & $19.16 \pm 0.82$ & $21.59 \pm 0.93$ & $39.17 \pm 1.68$ \\
              & 1--2 & 614 & $1.59 \pm 0.06$ & $32.63 \pm 0.93$ & $22.11 \pm 0.63$ & $20.65 \pm 0.59$ & $44.50 \pm 1.27$ \\
              & 2--3 & 656 & $2.46 \pm 0.10$ & $35.58 \pm 0.98$ & $26.91 \pm 0.74$ & $24.21 \pm 0.67$ & $50.76 \pm 1.40$ \\
All sample    & 3--4 & 601 & $3.50 \pm 0.14$ & $37.65 \pm 1.09$ & $26.73 \pm 0.77$ & $22.53 \pm 0.65$ & $51.38 \pm 1.48$ \\
              & 4--5 & 627 & $4.51 \pm 0.18$ & $36.22 \pm 1.02$ & $24.94 \pm 0.70$ & $21.16 \pm 0.60$ & $48.80 \pm 1.37$ \\
(N = 4, 829)  & 5--6 & 643 & $5.51 \pm 0.22$ & $38.79 \pm 1.08$ & $28.44 \pm 0.79$ & $22.26 \pm 0.62$ & $53.00 \pm 1.47$ \\
              & 6--7 & 696 & $6.50 \pm 0.25$ & $38.63 \pm 1.04$ & $27.06 \pm 0.73$ & $23.49 \pm 0.63$ & $52.69 \pm 1.42$ \\
              & 7--8 & 524 & $7.43 \pm 0.32$ & $40.62 \pm 1.26$ & $32.25 \pm 1.00$ & $24.35 \pm 0.75$ & $57.30 \pm 1.77$ \\
              & $>8$ & 197 & $8.60 \pm 0.61$ & $42.01 \pm 2.12$ & $29.40 \pm 1.48$ & $26.20 \pm 1.32$ & $57.58 \pm 2.90$ \\
\hline
              & 0--1 & 258 & $0.64 \pm 0.04$ & $23.68 \pm 1.04$ & $16.74 \pm 0.74$ & $14.51 \pm 0.64$ & $32.43 \pm 1.43$ \\
              & 1--2 & 595 & $1.59 \pm 0.07$ & $31.02 \pm 0.90$ & $19.82 \pm 0.58$ & $17.34 \pm 0.50$ & $40.69 \pm 1.18$ \\
              & 2--3 & 624 & $2.46 \pm 0.10$ & $32.55 \pm 0.92$ & $21.86 \pm 0.62$ & $18.53 \pm 0.52$ & $43.37 \pm 1.23$ \\
$TD/D \le 1$  & 3--4 & 569 & $3.50 \pm 0.15$ & $34.57 \pm 1.03$ & $21.41 \pm 0.64$ & $19.13 \pm 0.57$ & $44.94 \pm 1.34$ \\
              & 4--5 & 601 & $4.51 \pm 0.18$ & $34.53 \pm 1.00$ & $22.32 \pm 0.64$ & $18.14 \pm 0.52$ & $44.94 \pm 1.30$ \\
(N = 4, 578)  & 5--6 & 609 & $5.51 \pm 0.22$ & $35.69 \pm 1.02$ & $23.13 \pm 0.66$ & $18.55 \pm 0.53$ & $46.40 \pm 1.33$ \\
              & 6--7 & 650 & $6.49 \pm 0.25$ & $36.61 \pm 1.02$ & $23.13 \pm 0.64$ & $18.87 \pm 0.52$ & $47.24 \pm 1.31$ \\
              & 7--8 & 491 & $7.43 \pm 0.34$ & $38.15 \pm 1.22$ & $24.90 \pm 0.80$ & $19.31 \pm 0.62$ & $49.48 \pm 1.59$ \\
              & $>8$ & 181 & $8.60 \pm 0.64$ & $37.80 \pm 1.99$ & $25.06 \pm 1.32$ & $20.27 \pm 1.07$ & $49.68 \pm 2.62$ \\
\hline
                                  & 0--1 & 255 & $0.65 \pm 0.04$ & $25.11 \pm 1.11$ & $18.23 \pm 0.81$ & $13.05 \pm 0.58$ & $33.66 \pm 1.49$ \\
                                  & 1--2 & 593 & $1.59 \pm 0.07$ & $31.65 \pm 0.92$ & $21.28 \pm 0.62$ & $16.69 \pm 0.49$ & $41.63 \pm 1.21$ \\
                                  & 2--3 & 623 & $2.46 \pm 0.10$ & $34.34 \pm 0.97$ & $25.83 \pm 0.73$ & $17.52 \pm 0.50$ & $46.40 \pm 1.31$ \\
$Z_{\text{max}} \le 1\text{ kpc}$ & 3--4 & 564 & $3.50 \pm 0.15$ & $36.74 \pm 1.09$ & $25.24 \pm 0.75$ & $17.58 \pm 0.52$ & $47.92 \pm 1.42$ \\
                                  & 4--5 & 599 & $4.51 \pm 0.18$ & $35.26 \pm 1.02$ & $23.91 \pm 0.69$ & $17.27 \pm 0.50$ & $45.97 \pm 1.33$ \\
(N = 4, 537)                      & 5--6 & 606 & $5.51 \pm 0.22$ & $36.91 \pm 1.06$ & $25.13 \pm 0.72$ & $17.54 \pm 0.50$ & $47.97 \pm 1.38$ \\
                                  & 6--7 & 647 & $6.50 \pm 0.26$ & $37.79 \pm 1.05$ & $26.10 \pm 0.73$ & $17.73 \pm 0.49$ & $49.23 \pm 1.37$ \\
                                  & 7--8 & 479 & $7.43 \pm 0.34$ & $38.47 \pm 1.24$ & $28.74 \pm 0.93$ & $17.96 \pm 0.58$ & $51.27 \pm 1.65$ \\
                                  & $>8$ & 171 & $8.60 \pm 0.66$ & $37.33 \pm 2.02$ & $26.70 \pm 1.45$ & $18.07 \pm 0.98$ & $49.32 \pm 2.67$ \\
\hline\hline
\end{tabular}
\end{table*}

Figure~\ref{fig:age-dis} and Table~\ref{tab:age-kinematic} show the AVR for these three (sub)samples, where a clear increase in velocity dispersion with increasing total age is observed. For the complete sample, the total space velocity dispersion increases from $\sigma_{\rm tot} = 39.2$ km s$^{-1}$ for the youngest population ($<1$ Gyr) to 57.6 km s$^{-1}$ for stars older than 8 Gyr. A similar monotonic increase is calculated for the individual space velocity components, with the radial component exhibiting the strongest evolution, increasing from $\sigma_{\rm U} = 26.5$ to 42.0 km s$^{-1}$. The azimuthal component grows from $\sigma_{\rm V} = 19.2$ to 29.4 km s$^{-1}$, whereas the vertical space velocity component increases more gradually, from $\sigma_{\rm W} = 21.6$ to 26.2 km s$^{-1}$. These trends indicate that older WDs have experienced more prolonged dynamical heating through cumulative gravitational perturbations within the Galactic disk. 

To quantify the AVR, each velocity component was fitted with a power-law relation of the form
$\sigma_{\rm i} = A\times t^{B}$, where $t$ is the total stellar age in Gyr. For the complete sample, the best-fitting exponents are $B$ = 0.155 $\pm$ 0.015, 0.167 $\pm$ 0.033, 0.058 $\pm$ 0.029, and 0.137 ± 0.019 for the $\sigma_{\rm U}$, $\sigma_{\rm V}$, $\sigma_{\rm W}$, and $\sigma_{\rm tot}$, respectively (Table~\ref{tab:power_law}). The relatively large exponents found for the radial and azimuthal components demonstrate that in-plane stellar motions continue to be dynamically heated throughout the lifetime of the Galactic disk. In contrast, the substantially smaller exponent obtained for the vertical component suggests that vertical heating proceeds more slowly than radial heating, consistent with expectations for secular disk evolution. 

\begin{table}[ht]
\centering
\setlength{\tabcolsep}{3.5pt}
\caption{Best-fit parameters of the AVR, $\sigma = A \times t^{B}$, for the three WD samples. The fitted coefficients $A$ and $B$ and the corresponding correlation coefficients ($R^{2}$) are listed.}
\label{tab:power_law}
\begin{tabular}{ccccc}
\hline\hline
Sample & $\sigma$  & $A$ & $B$ & $R^2$ \\
       & (km~s$^{-1}$) &        &     &            \\
\hline
All sample & $\sigma_{\rm U}$ & $29.76 \pm 0.73$ & $0.155 \pm 0.015$ & $0.945$ \\
 & $\sigma_{\rm V}$ & $21.12 \pm 1.13$ & $0.167 \pm 0.033$ & $0.808$ \\
($N = 4,829$) & $\sigma_{\rm W}$ & $21.28 \pm 0.95$ & $0.058 \pm 0.029$ & $0.373$ \\
 & $\sigma_{\text{tot}}$ & $42.25 \pm 1.24$ & $0.137 \pm 0.019$ & $0.898$ \\
\hline
$TD/D\le 1$ & $\sigma_{\rm U}$ & $27.44 \pm 0.70$ & $0.160 \pm 0.016$ & $0.944$ \\
 & $\sigma_{\rm V}$ & $18.28 \pm 0.38$ & $0.143 \pm 0.013$ & $0.950$ \\
($N = 4,578$) & $\sigma_{\rm W}$ & $16.09 \pm 0.46$ & $0.099 \pm 0.018$ & $0.824$ \\
 & $\sigma_{\text{tot}}$ & $36.69 \pm 0.79$ & $0.145 \pm 0.013$ & $0.950$ \\
\hline
$Z_{\text{max}} \le 1\text{kpc}$ & $\sigma_{\rm U}$ & $28.99 \pm 0.94$ & $0.140 \pm 0.020$ & $0.887$ \\
 & $\sigma_{\rm V}$ & $20.40 \pm 0.92$ & $0.142 \pm 0.028$ & $0.803$ \\
($N = 4,537$) & $\sigma_{\rm W}$ & $15.06 \pm 0.52$ & $0.095 \pm 0.022$ & $0.748$ \\
 & $\sigma_{\text{tot}}$ & $38.53 \pm 1.26$ & $0.134 \pm 0.021$ & $0.875$ \\
\hline\hline
\end{tabular}
\end{table}

The same overall behavior is recovered in both thin-disk subsamples, although with systematically lower velocity dispersions owing to the reduced contribution from thick-disk and halo stars. In the kinematically selected thin-disk sample ($TD/D \leq 1$), the total space velocity dispersion increases from $\sigma_{\rm tot}$ = 32.4 to 49.7 km s$^{-1}$, while the orbital thin-disk sample ($Z_{\rm max} \leq 1$ kpc) exhibits a comparable increase from $\sigma_{\rm tot}$ = 33.7 to 49.3 km s$^{-1}$. The vertical space velocity dispersion remains consistently lower in the orbital sample, reflecting the explicit exclusion of stars reaching large distances from the Galactic plane. The power-law indices derived for both thin-disk samples ($B\approx0.10-0.16$) closely match those obtained for the full sample, confirming that the observed AVR is an intrinsic property of the thin-disk WD population rather than a consequence of contamination by older Galactic components. 

The observed increase in velocity dispersion with total age provides strong evidence that WDs retain the dynamical signatures of the secular heating history of the Galactic disk \citep[e.g.,][]{Rowell2019, Raddi2022}. The derived power-law exponents ($B\approx0.06-0.17$) are comparable to those reported for other Galactic stellar populations, indicating that WDs follow the same long-term dynamical evolution despite spanning a wide range of progenitor masses and evolutionary histories \citep[e.g.,][]{Holmberg2009, Ting2019}. 
The stronger age dependence of the radial velocity dispersion further supports the interpretation that scattering by giant molecular clouds, spiral arms, and the Galactic bar has played a dominant role in shaping the kinematic evolution of the local WD population \citep[e.g.,][]{Minchev2010, Grand2016}.

\subsection{Radial Migration and Orbital Blurring Across Galactic Populations}
We reconstruct the birth radii ($R_{\rm b}$) of each WD by integrating their Galactic orbits backward over their total stellar ages following the methodology of \citet{Akbaba2024}.  Because metallicity-based estimates are unavailable, we take each star’s Galactocentric radius at that lookback time as an estimate of its birth radius. This assumes that the adopted Galactic potential approximates the star’s past orbit; the integration cannot independently recover changes in angular momentum caused by past interactions. We propagate uncertainties in the astrometric parameters, distances, proper motions, and radial velocities through the orbital integrations to estimate the uncertainties in $R_{\rm b}$. We then compare $R_{\rm b}$ with the present guiding radius ($R_{\rm g}$) and Galactocentric radius ($R_{\rm gc}$). Interpretations of the resulting displacements and migration directions depend on the adopted birth-radius estimate. For comparison, we overplot the age-dependent radial-migration relations of \citet{Frankel2018, Frankel2020} as presented in Figure~9 of \citet{Netopil2022}. These curves are used only as a reference; no additional transformations of the dynamical quantities reported by \citet{Frankel2018, Frankel2020} were applied. WDs lying below or above these curves exhibit radial displacements that are smaller or larger, respectively, than the corresponding reference values at a given age. However, these differences should not be interpreted directly as evidence for weaker or stronger dynamical interactions, because the inferred displacements also depend on individual orbital histories, birth radii, age uncertainties, and the assumptions underlying the migration models.

\begin{figure*}[ht]
\centering
\includegraphics[width=2\columnwidth]{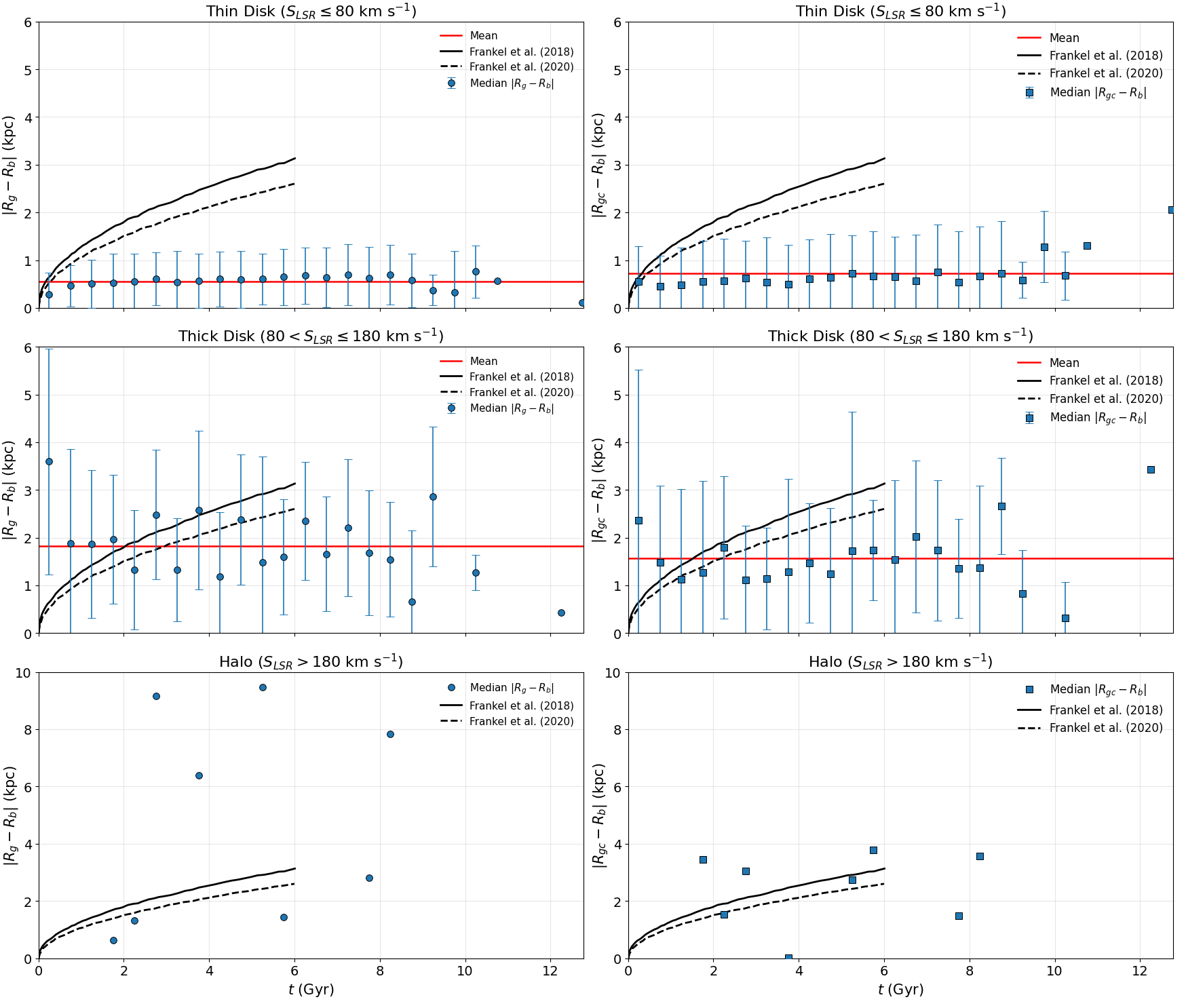}
\caption {Radial displacement versus total stellar age for thin-disk (top), thick-disk (middle), and halo (bottom) WDs. The left and right columns show $|R_{\rm g}-R_{\rm b}|$ and $|R_{\rm gc}-R_{\rm b}|$, respectively. The populations are separated at $S_{\rm LSR}=80$ and 180~km~s$^{-1}$. Blue symbols show the median displacement in 0.5~Gyr age bins; error bars show the standard deviation within each bin. Red horizontal lines mark the mean displacement for each disk subsample.
The solid and dashed black curves are the reference relations of \citet{Frankel2018} and \citet{Frankel2020}, respectively.} 
\label{fig:Pop-Chu-Blu}
\end {figure*}

Figure~\ref{fig:Pop-Chu-Blu} presents $|R_{\rm g}-R_{\rm b}|$ and $|R_{\rm gc}-R_{\rm b}|$ for the thin-disk, thick-disk, and halo WDs in separate panels. For this analysis, we separate the populations using the $S_{\rm LSR}$ values, following the kinematic criteria of \citet{Nissen2004}. This velocity-based assignment is distinct from the probability-based classification used elsewhere in the paper. The difference $|R_{\rm g}-R_{\rm b}|$ compares the present $R_{\rm g}$ with the inferred $R_{\rm b}$ and thus serves as a proxy for radial migration associated with changes in angular momentum. By contrast, $|R_{\rm gc}-R_{\rm b}|$ uses the present-day $R_{\rm gc}$; it includes both the displacement of the guiding radius and the star's current epicyclic excursion about that radius. The two quantities therefore describe related, but non-additive, aspects of radial redistribution.

We show their age dependence using the median displacement in 0.5~Gyr bins. The error bars show the standard deviation within each bin, while the red lines mark the mean displacement of each disk subsample. The thin-disk WDs have mean $|R_{\rm g}-R_{\rm b}|$ and $|R_{\rm gc}-R_{\rm b}|$ values of $0.55\pm 0.58$ and $0.73\pm 0.83$~kpc, respectively. The median $|R_{\rm g}-R_{\rm b}|$ increases modestly from the youngest bins to ages of about 7--8~Gyr, although the sequence is not monotonic. The median $|R_{\rm gc}-R_{\rm b}|$ shows no clear age trend. At ages of a few Gyr and older, most thin-disk WDs fall below the reference relations. The thick-disk WDs show larger mean displacements, $1.82\pm 1.42$~kpc in $|R_{\rm g}-R_{\rm b}|$ and $1.57\pm 1.55$~kpc in $|R_{\rm gc}-R_{\rm b}|$. Their medians vary substantially with age and do not establish a monotonic trend. Only 11 halo WDs have total stellar ages. Their displacements span a wide range, but the small sample does not support a characteristic migration distance or a reliable age trend.

Comparison of the $R_{\rm g}$ and $R_{\rm gc}$ displacements shows that the characteristic values inferred from the two measures are of comparable magnitude in both disk populations. The displacement relative to $R_{\rm g}$ is therefore substantial compared with that relative to the present-day position, under our adopted estimate of $R_{\rm b}$. However, because $|R_{\rm g}-R_{\rm b}|$ and $|R_{\rm gc}-R_{\rm b}|$ do not constitute a strict additive decomposition of radial migration and orbital blurring, we do not assign separate fractional contributions to these two processes.

\begin{figure}[ht]
\centering
\includegraphics[width=1\columnwidth]{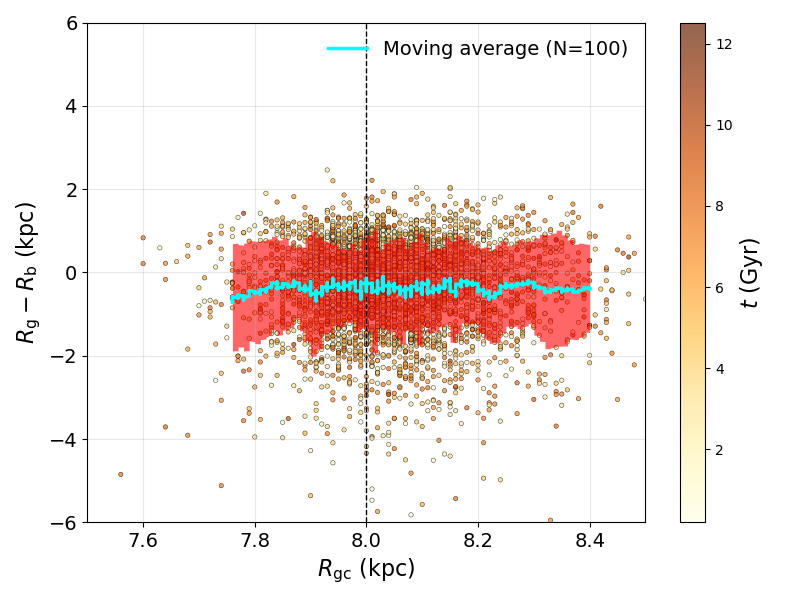}
\caption{Radial displacement ($|R_{\rm g}-R_{\rm b}|$ vs the present-day Galactocentric radius ($R_{\rm gc}$) for the WDs in our sample. 
Each object is color-coded according to its total age. The dashed line marks the Galactocentric distance of the Sun, the cyan curve represents the moving average, and the red shaded region shows the $1\sigma$ interval.} 
\label{fig:Directions}
\end {figure}

We examine the direction of radial migration by comparing the reconstructed birth radius $R_{\rm b}$ with the present-day guiding radius. WDs with $R_{\rm g} - R_{\rm b} > 0$ kpc are classified as outward migrators, whereas those with $R_{\rm g} - R_{\rm b} < 0$ kpc are identified as inward migrators. Figure~\ref{fig:Directions} illustrates the distribution of migration direction as a function of $R_{\rm gc}$, with stellar ages indicated by the color scale. Among the 4,829 WDs with reliable total ages, 1,933 stars (40\%) migrated outward from the inner Galactic disk, while 2,896 stars (60\%) migrated inward from larger Galactocentric radii. The outward-migrating population exhibits a mean radial displacement of +0.57 kpc and a mean age of 4.3 Gyr, whereas the inward migrators show an average displacement of $-1.00$ kpc and a mean age of 4.5 Gyr. The nearly identical mean ages of the two groups indicate that the asymmetry is not driven by age.

Although inward migrators outnumber outward migrators in the present sample, this imbalance is not evidence for a globally preferred inward migration pattern. Radial-migration models generally permit migration in both directions. The diffusion model of \citet{Frankel2018} can produce an overall outward broadening of the stellar disk, whereas the updated model of \citet{Frankel2020} includes a small inward drift term to approximately conserve angular momentum, making inward migration somewhat more probable while retaining substantial migration in both directions. Thus, the relative numbers of inward and outward migrators are model-dependent rather than a universal prediction of radial-migration theory. Given the uneven sky coverage of the SDSS observations, our sample does not provide a spatially complete census of the Galactic disk. Approximately 63\% of the objects in our sample lie inside the solar circle, whereas 37\% are located outside of it, closely matching the relative fractions of inward and outward migrators. Hence, the observed asymmetry in the number of inward and outward migrators is likely due to the observational biases in the SDSS. Future analyses using more spatially complete WD samples will be essential for more stringent constraints on the global migration pattern of the Milky Way disk. 

\begin{table*}
\centering
\scriptsize
\caption{Galactocentric mean velocity components and space-velocity dispersions of the WD subsamples used in the Jeans analysis.}
\label{tab:galactic_velocity}
\begin{tabular}{llccccccccccc}
\hline
Sample & Population & $N$ & $\langle V_R\rangle$ & $\langle V_\phi\rangle$ & $\langle V_Z\rangle$ & $\sigma_{V_{\rm R}}$ & $\sigma_{V_\phi}$ & $\sigma_{V_Z}$ & $\partial\sigma_{{\rm V_{\rm Z}}}/\partial Z$ & $K_{\rm Z}$ & $h_{\rm R}$ & $h_{\rm Z}$\\
\cline{4-9}
       &            &     & \multicolumn{6}{c}{(km\,s$^{-1}$)} & (km~s$^{-1}$~kpc$^{-1}$) & ((km~s$^{-1}$)$^2$~pc$^{-1}$) & (kpc) & (pc) \\
\hline
$|Z|<800$ pc              & Thin disk  & 5,825 & 1.61 & -208.01  & 1.52  & 38.19 & 27.66 & 24.33 & 13.47 & 1.93 & 3.49 & 229 \\
$TD/D\leq1$               & Thin disk  & 5,468 & 1.18 & -209.70 & -1.59 & 35.23 & 22.66 & 18.93 & 13.47 & 1.93 & 3.29 & 147 \\
$0<Z_{\rm max}\leq1$ kpc  & Thin disk  & 5,408 & 1.47 & -206.97 & -1.58 & 36.39 & 25.33 & 17.73 & 21.39 & 1.93 & 3.22 & 123 \\
$1<Z_{\rm max}\leq4$ kpc  & Thick disk & ~~412 & 3.89 & -165.13 & 21.70 & 54.69 & 43.19 & 57.74 & 24.39 & 2.82 & 2.97 & 865 \\
\hline
\end{tabular}
\end{table*}

\subsection{Disk Scale Height and Scale Length}

We use three different subsamples to constrain the radial and vertical scale lengths of the Galactic disk and to investigate how the adopted population affects the inferred Galactic structure. The subsamples are selected based on (i) the present-day vertical distance from the Galactic plane ($|Z|<800$ pc), (ii) the kinematic population probabilities of \citet{Bensby2003, Bensby2005}, and (iii) the maximum orbital height above the Galactic plane ($Z_{\rm max}$). 

Table~\ref{tab:galactic_velocity}
presents the Galactocentric cylindrical velocity components and their corresponding velocity dispersions for each subsample. The derived kinematic properties clearly
distinguish the thin- and thick-disk populations. The thin-disk subsamples are characterized by mean radial and vertical velocities close to zero together with mean rotational velocities $\langle V_\phi\rangle$ between $-207$ and $-210$ km~s$^{-1}$, indicating that these stars participate in the regular rotation of the Galactic disk. In contrast, the thick-disk sample exhibits a substantially lower rotational velocity, $\langle V_{\phi}\rangle\approx-165$ km~s$^{-1}$, and larger velocity dispersions, particularly in the vertical direction ($\sigma_{{\rm V_{\rm Z}}}=57.7$ km~s$^{-1}$).

\begin{figure}[t]
\centering
\includegraphics[width=1\columnwidth]{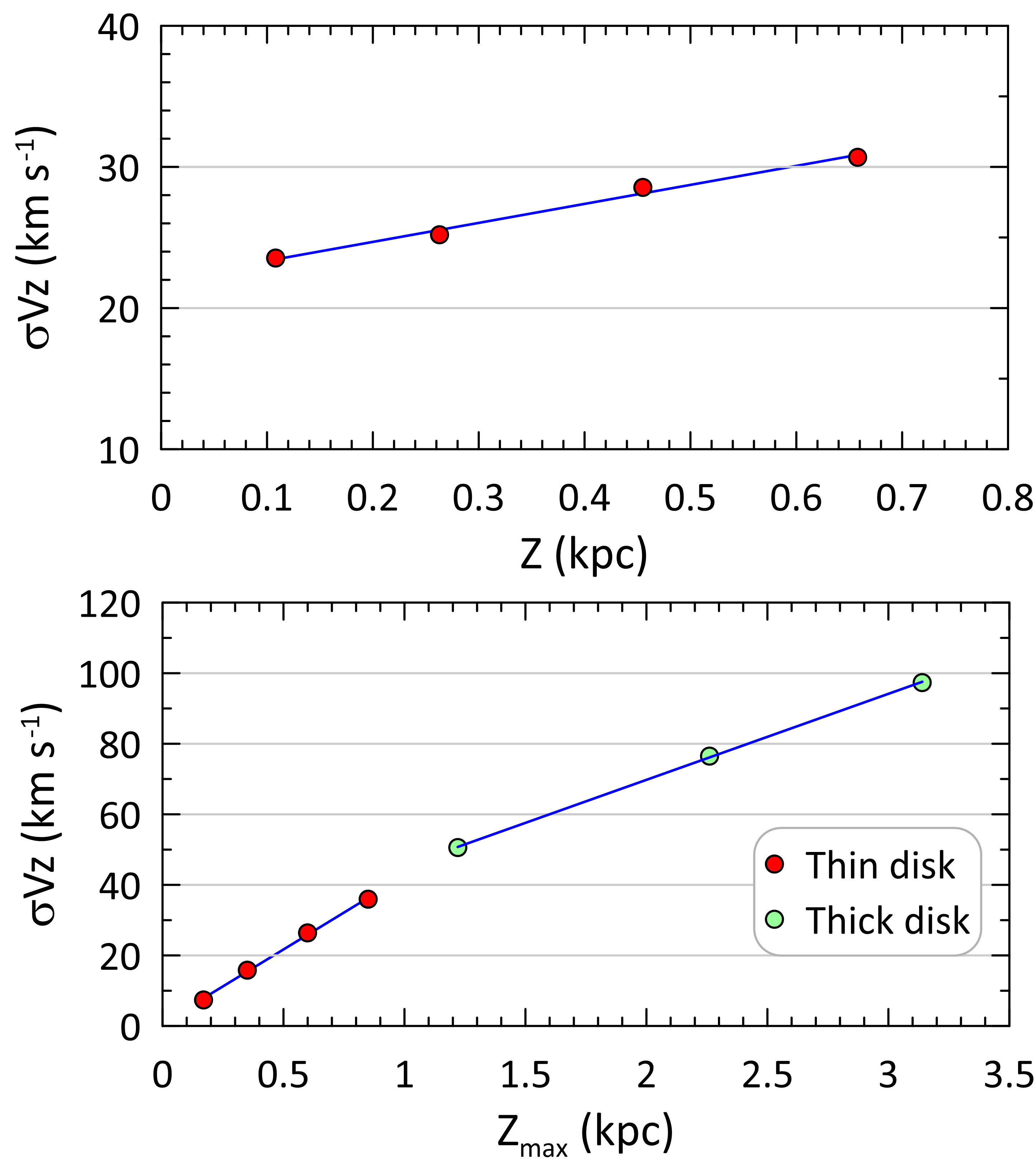}
\caption {Velocity dispersion as a function of distance from the Galactic plane. The upper panel shows the present-day vertical distances ($Z$) of the WDs from the Galactic plane, while the lower panel shows their maximum vertical distances ($Z_{\rm max}$) reached along their Galactic orbits.} 
\label{fig:Velocity-dispersion_gradient}
\end {figure}

The vertical scale heights were determined using Equation~(\ref{eqn:vertical_jeans}), which combines the measured vertical velocity dispersions with their vertical gradients and the corresponding vertical Galactic force ($K_{\rm Z}$). The velocity-dispersion gradients were estimated from the variation of $\sigma_{{\rm V}_{\rm Z}}$ with both the present-day vertical distance and the maximum orbital height  (Figure~\ref{fig:Velocity-dispersion_gradient}), while the adopted $K_{\rm Z}$ values were obtained from the $K_{\rm Z}(Z)$ relation presented in Figure~14 of \citet{Kuijken1989}. Representative $K_{\rm Z}$ values were selected according to the vertical ranges adopted for the two disk populations, $0<|Z|~{\rm (kpc)}\leq1$ for the thin disk and $1<|Z|~{\rm (kpc)}\leq4$ for the thick disk. For the thin disk, we adopted $K_{\rm Z}=1.93~({\rm km~s}^{-1})^2~{\rm pc}^{-1}$ for the vertical distance interval $|Z|\leq1.1$ kpc \citep[see also,][]{Kuijken1989, Kordopatis2011}. For example, for the thick disk, the midpoint of the adopted vertical interval, $|Z|=2.5$~kpc, corresponds to $K_{\rm Z}=2.82~({\rm km~s}^{-1})^2~{\rm pc}^{-1}$. Thus, a single representative $K_{\rm Z}$ value was adopted for each vertical interval in the scale-height calculation.

This $K_{\rm Z}$ prescription was adopted to follow the Jeans-equation implementation used in previous scale-height analyses \citep[e.g.,][]{Kordopatis2011}, rather than being derived from the \texttt{MWPotential2014} model employed for the orbit integrations. We note that this choice introduces an explicit dependence of the inferred $h_{\rm Z}$ values on the adopted vertical-force model. This procedure allows each Galactic population to be characterized using dynamical parameters appropriate for its typical orbital distribution. The calculated velocity gradients ($\partial\sigma_{{\rm V_{\rm Z}}}/\partial Z$) and $K_{\rm Z}$ values for the thin- and thick-disk components are included in Table~\ref{tab:galactic_velocity}. 

We emphasize that the scale heights inferred from Equation~(\ref{eqn:vertical_jeans}) are conditional on the adopted Galactic vertical force. In particular, the $K_{\rm Z}/\sigma_{V_{\rm Z}}^2$ term provides a substantial contribution to $1/h_{\rm Z}$ for our sample, so that the resulting scale heights should not be interpreted as purely model-independent measurements. Their values also depend on the adopted sample selection through both $\sigma_{V_{\rm Z}}$ and its vertical gradient.

For the thin-disk sample selected solely according to its present-day spatial distribution ($|Z|<800$ pc), the scale height and radial scale length are $h_{\rm Z}=229$ pc and $h_{\rm R}=3.49$ kpc, respectively. Applying the kinematic population criterion ($TD/D\le1$) reduces the scale height to $147$ pc while yielding a comparable radial scale length of $3.29$ kpc. The smallest thin-disk scale height is obtained for the orbital selection based on $Z_{\rm max}$, resulting in $h_{\rm Z}=123$ pc and $h_{\rm R}=3.22$ kpc. In contrast, the thick-disk ($1<Z_{\rm max}~{\rm (kpc)} \leq4$) population exhibits a substantially larger scale height of $h_{\rm Z}=865$ pc, whereas its radial scale length remains comparatively similar ($h_{\rm R}=2.97$ kpc). The derived radial scale lengths and vertical scale heights for all populations are included in Table~\ref{tab:galactic_velocity}. The relatively small variation in the radial scale length indicates that the principal structural difference between the two Galactic disk components lies in their vertical distribution rather than their radial extent. The orbital classification based on $Z_{\rm max}$ yields the smallest thin-disk scale height, suggesting that orbital parameters provide a more effective separation of thin- and thick-disk stars than either present-day positions or purely kinematic probability criteria. 

\section{Discussion}
\label{sec:summary-and-discussion}

\subsection{Thin-Disk, Thick-Disk, and Halo Velocity Dispersions}

\begin{table*}
\centering
\caption{Thin-disk, Thick-disk, and halo velocity dispersions.}
\label{tab:comparison}
\begin{tabular}{lcccccl}
\hline
Source & $N$ & $\sigma_{\rm U}$ & $\sigma_{\rm V}$ & $\sigma_{\rm W}$ & Population & Object \\
\cline{3-5}
 & & \multicolumn{3}{c}{(km\,s$^{-1}$)} & & \\
\hline

\multicolumn{7}{c}{Thin Disk}\\
\hline
\citet{Pauli2003}            & 361  & 34          & 24          & 18          & Thin disk & White dwarfs \\
\citet{Soubiran2003}         & ---  & $39\pm2$    & $20\pm2$    & $20\pm1$    & Thin disk & Solar neighborhood \\
\citet{Bilir2005}            & 129  & $36\pm7$    & $26\pm7$    & $19\pm5$    & Thin disk & W Ursae Majoris  \\
\citet{Ak2010}               & 131  & $29\pm5$    & $21\pm4$    & $17\pm4$    & Thin disk & Cataclysmic variables \\
\citet{Bensby2014}           & 387  & 35          & 20          & 16          & Thin disk & Solar neighborhood \\
\citet{Ak2015}               & 199  & $34\pm4$    & $28\pm4$    & $18\pm4$    & Thin disk & Cataclysmic variables \\
\citet{Elsanhoury2015}       & 141  & 39          & 27          & 32          & Thin disk & White dwarfs \\
\citet{Canbay2025}           & 354  & $34\pm3$    & $24\pm3$    & $18\pm2$    & Thin disk & Cataclysmic variables \\
This study ($S_{\rm LSR}\le80$ km\,s$^{-1}$) & 5,304 & $32\pm3$  & $22\pm3$ & $20\pm3$ & Thin disk & White dwarfs \\
This study ($TD/D\le1$)      & 5,468 & $35\pm3$    & $23\pm3$    & $19\pm3$    & Thin disk & White dwarfs \\
This study ($Z_{\rm max}\le1$ kpc) & 5,408 & $36\pm3$ & $25\pm3$ & $18\pm3$ & Thin disk & White dwarfs \\
\hline
\multicolumn{7}{c}{Thick Disk}\\
\hline
\citet{Chiba2000}                                      & ---  & $46\pm4$    & $50\pm4$    & $35\pm3$    & Thick disk & Solar neighborhood \\
\citet{Pauli2003}                                      & 27   & 79          & 36          & 46          & Thick disk & White dwarfs \\
\citet{Soubiran2003}                                   & ---  & $63\pm6$    & $39\pm4$    & $39\pm4$    & Thick disk & Solar neighborhood \\
\citet{Bensby2014}                                     & 203  & 67          & 38          & 35          & Thick disk & Solar neighborhood \\
\citet{Canbay2025}                                     & 13   & $56\pm2$    & $50\pm4$    & $34\pm2$    & Thick disk & Cataclysmic variables \\
This study ($80<S_{\rm LSR}~({\rm km~s}^{-1})\le180$)  & 516  & $75\pm3$    & $41\pm3$    & $46\pm3$    & Thick disk & White dwarfs \\
This study ($TD/D>10$ and $TD/H> 1$)                   & 186  & $71\pm 4$   & $47\pm 2$   & $67\pm 3$   & Thick disk & White dwarfs \\
This study ($1<Z_{\rm max}\le4$ kpc)                   & 412  & $54\pm3$    & $43\pm3$    & $58\pm3$    & Thick disk & White dwarfs \\
\hline
\multicolumn{7}{c}{Halo}\\
\hline
\citet{Chiba2000}                             & ---  & $141\pm11$  & $106\pm9$   & $94\pm8$    & Halo & Solar neighborhood \\
\citet{Pauli2003}                             & 7    & 138         & 95          & 47          & Halo & White dwarfs \\
\citet{Ak2010}                                & 10   & $78\pm4$    & $63\pm5$    & $48\pm5$    & Halo & Cataclysmic variables \\
\citet{Ak2015}                                & 17   & $59\pm3$    & $50\pm3$    & $66\pm4$    & Halo & Cataclysmic variables \\
\citet{Bensby2014}                            & 36   & 160         & 90          & 90          & Halo & Solar neighborhood \\
\citet{Canbay2025}                            & 18   & $78\pm2$    & $75\pm2$    & $62\pm3$    & Halo & Cataclysmic variables \\
This study ($S_{\rm LSR}>180$ km\,s$^{-1}$)   & 17   & $132\pm23$  & $93\pm16$   & $120\pm21$  & Halo & White dwarfs \\
This study ($TD/D>10$ and $TD/H\leq 1$)       & 16   & $123\pm22$  & $103\pm19$  & $137\pm25$  & Halo & White dwarfs \\
This study ($Z_{\rm max}>4$ kpc)              & 17   & $91\pm16$   & $100\pm18$  & $151\pm27$  & Halo & White dwarfs \\

\hline
\end{tabular}
\end{table*}

Table~\ref{tab:comparison} presents the velocity dispersions for the thin disk, thick disk, and halo WDs in our sample along with representative values reported in the literature. Thin-disk WDs exhibit remarkably consistent velocity dispersions regardless of the adopted population criteria used ($S_{\rm tot}$, $TD/D$, or $Z_{\rm max}$). The derived values range from $(\sigma_{\rm U},\sigma_{\rm V},\sigma_{\rm W})=(32\pm3,\,22\pm3,\,20\pm3)$ km~s$^{-1}$ for the $S_{\rm tot}$ selection to $(36\pm3,\,25\pm3,\,18\pm3)$ km~s$^{-1}$ for the $Z_{\rm max}$ selection. These measurements are in excellent agreement with previous studies of WDs \citep{Pauli2003,  Elsanhoury2015}, nearby thin-disk stars \citep{Soubiran2003, Bensby2014}, W UMa-type variables \citep{Bilir2005}, and cataclysmic variables \citep{Ak2010, Ak2015, Canbay2025}. The close agreement between these independent stellar populations shows that the Gaia-based WD sample reliably reproduces the kinematic properties of the thin disk.

As expected, the thick-disk population exhibits substantially larger velocity dispersions. The largest radial velocity dispersion is obtained from the $S_{\rm tot}$ criterion, reaching $\sigma_{\rm U}=75\pm3$ km~s$^{-1}$, which is consistent with $\sigma_{\rm U}=79$ km~s$^{-1}$ reported for the SPY WD sample by \citet{Pauli2003}. The corresponding $\sigma_{\rm V}$ and $\sigma_{\rm W}$ values are likewise consistent with measurements for nearby thick-disk stars reported by \citet{Chiba2000}, \citet{Soubiran2003}, and \citet{Bensby2014}. The lower $\sigma_{\rm U}$ values derived from the probability-based and orbital classification methods are expected, since they use the complete kinematic or orbital properties rather than just the total space velocities, thereby reducing the contribution of objects with exceptionally large radial motions.

The halo subsamples display the largest velocity dispersions, reflecting the dynamically hottest Galactic population, though our results suffer from small-number statistics. The halo WD sample identified through the total space velocity displays velocity dispersions of $(\sigma_{\rm U},\sigma_{\rm V},\sigma_{\rm W})=(132\pm23,\,93\pm16,\,120\pm21)$ km~s$^{-1}$, which closely match the values reported by \citet{Chiba2000} and \citet{Pauli2003} in $\sigma_{\rm U}$ and $\sigma_{\rm V}$. The lower dispersions obtained from the probability-based classification most likely arise because of contamination from thick disk objects. 

All three classification methods recover the same fundamental kinematic sequence, namely a systematic increase in the velocity dispersions from the thin disk to the thick disk and finally to the halo. This behavior is fully consistent with the long-established picture of secular disk heating \citep{Wielen1977, Nordstrom2004, Aumer2009}, in which repeated gravitational interactions with molecular clouds, spiral arms, the Galactic bar, and external perturbations progressively increase stellar random motions over time \citep{Binney2008, Sellwood2014, Bland-Hawthorn2016}. 

\subsection{The Guiding Radius Distributions}

\begin{figure}[ht]
\centering
\includegraphics[width=1\columnwidth]{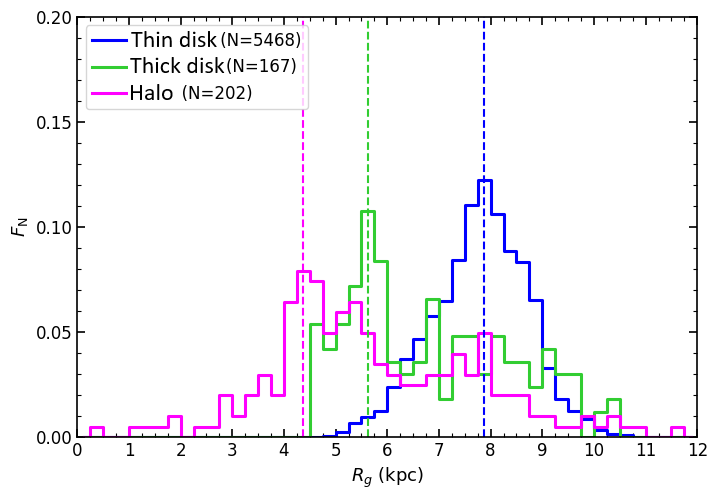}
\caption{Normalized histograms of the $R_{\rm g}$ distributions for the Galactic populations identified in this study.} 
\label{fig:Rg_pop}
\end {figure}

The guiding-radius distributions provide a useful dynamical comparison among the Galactic stellar populations because $R_{\rm g}$ is closely related to orbital angular momentum. The top panel in Figure~\ref{fig:Rg_pop} presents our results on the normalized $R_{\rm g}$ distributions for the thin-disk, thick-disk, and the halo. The distribution peaks at $R_{\rm g}=7.88$ kpc for the thin disk, $5.62$ kpc for the thick disk, and $4.38$ kpc for the halo, revealing the expected systematic decrease in characteristic orbital radius from dynamically cold to dynamically hot Galactic populations.

For comparison, we use the quantitative guiding-radius results reported by \citet{Zubiaur2024}. Their sample is based on the Galactic population classification of \citet{Torres2019}, who used a Random Forest algorithm trained on synthetic thin-disk, thick-disk, and halo populations generated through population-synthesis modeling. Thus, the population assignments adopted by \citet{Zubiaur2024} are probabilistic rather than being based on fixed kinematic thresholds. Despite the different classification methodology from that adopted in the present work, the resulting $R_{\rm g}$ distributions show broadly similar characteristic trends.  Specifically, \citet{Zubiaur2024} report mean guiding radii of approximately 7.9, 6.7, and 4.2~kpc for the thin-disk, thick-disk, and halo populations, respectively, providing a quantitative comparison with the values derived in the present study.

The nearly identical peak positions obtained for the thin-disk and halo populations demonstrate that the characteristic angular momenta of these populations are recovered consistently despite the use of different datasets and classification techniques. The modest offset of approximately 1 kpc for the thick disk is likely due to differences in sample selection, survey volume, and population assignment. 

\subsection{Galactic Structural Parameters}

We used three independent classification schemes ($S_{\rm tot}$, $TD/D$, and $Z_{\rm max}$) to constrain the relative fractions of the Galactic populations. For the full sample of 5,837 WDs, the three classification methods yield thin-disk fractions of 91–94\%, while the inferred thick-disk fraction varies more substantially between methods and the halo fraction remains below 1\%. These fractions are broadly consistent with Galactic population studies based on nearby field stars and large photometric surveys \citep[e.g.,][]{Robin2003, Karaali2003, Karaali2004, Karaali2007, Bilir2006a, Bilir2006b, Bilir2006c, Bilir2008, Ak2007, Juric2008,  Iyisan2025}, confirming that the local WD population is overwhelmingly dominated by the thin disk.

The structural parameters were derived for three independently selected thin-disk samples and one thick-disk sample. The thin-disk scale height decreases systematically from $h_{\rm Z}=229$ pc for the purely spatial sample ($|Z|<800$ pc) to $147$ pc for the kinematically selected sample ($TD/D\le1$), reaching its minimum value of $123$ pc for the orbitally selected sample ($Z_{\rm max}\le1$ kpc). In contrast, the radial scale length remains comparatively stable, varying between 3.22 and 3.49 kpc. 

The derived structural parameters compare favorably with previous Galactic structure studies. Although the exact values depend on the adopted stellar tracers and selection criteria, virtually all investigations find a thin disk characterized by a relatively small vertical scale height and a significantly thicker disk component extending several hundred parsecs above the Galactic plane \citep{Robin2003, Juric2008, Bovy2017}. The comparatively small thin-disk scale height in this study likely reflects the orbital selection procedure, which efficiently removes stars with large vertical excursions, along with the relatively young progenitor population represented by many of the WDs. Conversely, the thick-disk scale height of 865 pc agrees remarkably well with the $\sim900$ pc value reported by \citet{Juric2008}, supporting the robustness of the present analysis. The radial scale lengths derived in this study are somewhat larger than those reported for main-sequence and red-clump stars \citep{Bilir2006a, Yaz2015}. Such differences are expected because radial scale lengths depend on the adopted Galactic potential, sample selection, and the stellar tracer population. 

\subsection{Age--Velocity Dispersion Relation}

\begin{figure}[ht]
\centering
\includegraphics[width=1\columnwidth]{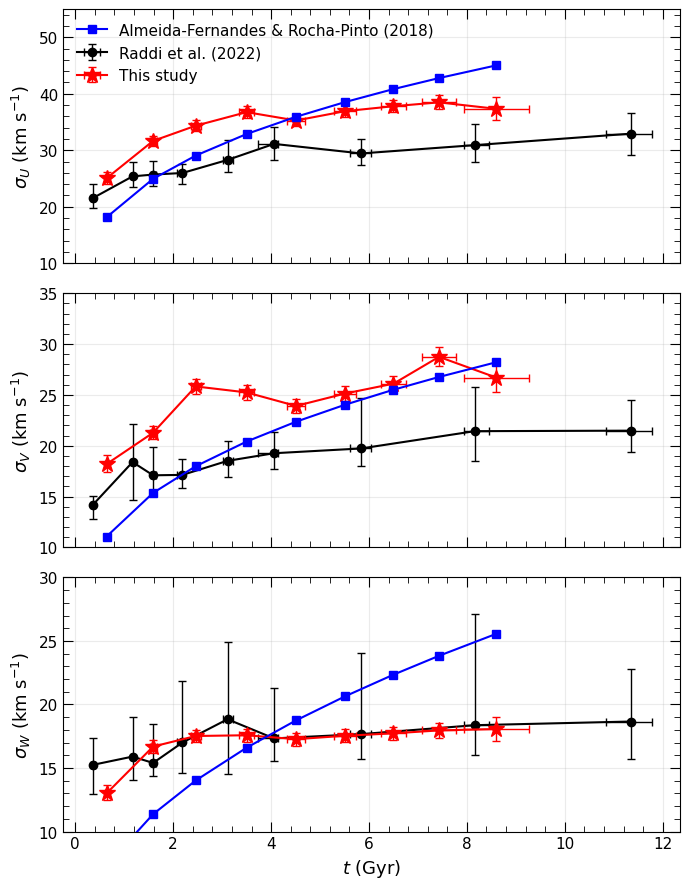}
\caption{Age-velocity dispersion relations found in this study (red) versus two other studies in the literature (black and blue points).}
\label{fig:literature}
\end {figure}

The AVR provides one of the strongest observational constraints on the secular evolution of the Galactic disk. Stars are born with relatively small random motions that gradually increase through gravitational interactions with giant molecular clouds, transient spiral arms, the Galactic bar, and satellite perturbations \citep{Spitzer1951, Spitzer1953, Wielen1977, Sellwood2002, Aumer2009}. Consequently, older stellar populations exhibit systematically larger velocity dispersions than younger populations, making the AVR a powerful tracer of the dynamical heating history of the Milky Way.

For our sample of WDs with reliable total ages, all three space velocity components exhibit a clear increase in velocity dispersion with increasing stellar age. The same trend is recovered for both the full sample and the thin-disk subsamples selected using the $TD/D\le1$ and $Z_{\rm max}\le1$ kpc criteria, indicating that the observed AVR is not an artifact of the adopted population selection. The correlation becomes stronger for the dynamically homogeneous thin-disk samples, where contamination from thick-disk and halo stars is minimized, demonstrating that the secular dynamical heating of the Galactic disk is robustly detected in the WD population.

Figure~\ref{fig:literature} compares our results with those obtained from the WD sample of \citet{Raddi2022} and the main-sequence star sample of \citet{Almeida-Fernandes2018}. The overall increase in $\sigma_{\rm U}$, $\sigma_{\rm V}$, and $\sigma_{\rm W}$ with age is consistent among all three investigations, confirming the universal nature of the AVR. Our sample exhibits systematically larger $\sigma_{\rm U}$ and $\sigma_{\rm V}$ values than those reported by \citet{Raddi2022}, though they are consistent with \citet{Almeida-Fernandes2018} for ages older than 4~Gyr. 

\citet{Arseneau25} found systematic velocity differences
of order 5-15 km s$^{-1}$ between the SDSS spectra and higher resolution SPY data. They found that the SDSS radial velocities tend to be higher
than expected from the gravitational redshifts and provided corrections for the resolution-induced redshifts. However, repeating our entire analysis
after including these corrections did not noticeably change the velocity dispersions shown in Figure~\ref{fig:literature}. In any case, given the errors, the differences between our results and those of \citet{Raddi2022} are not statistically significant. Furthermore,
the vertical velocity dispersion, $\sigma_{\rm W}$, agrees remarkably well with \citet{Raddi2022} measurements over the entire age range. Beyond $\sim4$ Gyr, the three independent studies converge toward nearly identical velocity dispersions, suggesting that the long-term dynamical evolution of the Galactic disk dominates over differences in stellar tracers and analysis techniques.

\subsection{Radial Migration and Orbital Blurring}

To study the impact of radial migration and orbital blurring on our WD sample, we divide the sample into thin-disk, thick-disk, and halo groups using the total space velocity criteria of \citet{Nissen2004}. This is the same kinematic grouping used in Figure 7 and differs from the probability-based classification applied elsewhere in the paper. The inferred birth radii are based on backward orbit integrations and therefore depend on the adopted Galactic potential. The mean $|R_{\rm g}-R_{\rm b}|$ and $|R_{\rm gc}-R_{\rm b}|$ values are $0.55\pm 0.58$ and $0.73\pm 0.83$ kpc for the thin disk and $1.82\pm 1.42$ and $1.57\pm 1.55$ kpc for the thick disk, respectively. Thus, both displacement measures are larger in the thick-disk group. The binned thin-disk medians show a modest, non-monotonic increase with age in $|R_{\rm g}-R_{\rm b}|$, whereas the thick-disk medians show no clear age trend. The substantial scatter cautions against interpreting individual displacements as direct measures of past dynamical interactions. Only 11 halo WDs have total-age estimates, preventing a reliable determination of an age trend or characteristic displacement for that group. The two disk displacement measures have comparable magnitudes, suggesting that the difference between the present guiding radius and the inferred birth radius is relevant to the radial distribution of both populations. However, because these quantities do not provide an additive decomposition of radial migration and orbital blurring, we do not assign separate fractional contributions to the two processes.

Although inward migrators outnumber outward migrators, the uneven spatial coverage of the SDSS sample may affect their observed ratio. The present data therefore do not establish a preferred migration direction for the Galactic disk. A spatially more complete WD sample is required to constrain the overall balance between inward and outward migration.

The radial-displacement trends found for the WD sample are broadly consistent with studies based on other stellar tracers. Bayesian reconstructions of stellar birth radii using APOGEE red giants show that many solar-neighborhood stars formed several kiloparsecs from their present locations \citep{Frankel2018, Frankel2020}. Evidence for radial redistribution is also seen in red-clump stars \citep{Anders2017}, FGK disk stars \citep{Sharma2021}, open clusters \citep{Netopil2022, Canbay2026, Cinar2026a, Cinar2026b, Yontan2026}, and APOGEE DR17 samples of main-sequence turnoff and subgiant branch stars \citep{Queiroz2023}. Taken together, these studies demonstrate that radial redistribution is an important aspect of Galactic disk evolution across a wide range of stellar tracers. 

\section{Conclusions and Future Prospects}
\label{sec:con}

We have presented a comprehensive kinematic and dynamical analysis of a sample of 5,837 DA WDs constructed by combining SDSS DR19 spectroscopy \citep{Crumpler2025} with Gaia DR3 astrometry \citep{Gaia2023}. High-precision Gaia astrometric measurements and homogeneous atmospheric parameters from the SDSS provide one of the largest and most reliable WD samples currently available for investigating the dynamical evolution of the Galactic disk. 

The kinematic analysis shows that the sample is dominated by thin-disk WDs, while smaller fractions belong to the thick disk and halo. Independent population classifications based on total space velocity, the kinematic probability formalism of \citet{Bensby2003, Bensby2005}, and Galactic orbital parameters yield consistent results, demonstrating that the different approaches identify similar Galactic populations. In all classification schemes, velocity dispersions increase systematically from the thin disk to the halo, accompanied by a progressively larger rotational lag and increasing orbital eccentricities. These trends indicate the transition from dynamically cold stellar populations on nearly circular Galactic orbits to dynamically hotter populations characterized by larger random motions and more eccentric trajectories.

The availability of reliable total stellar ages allows the AVR to be examined over a wide range of Galactic ages. The velocity dispersions of all three velocity components increase systematically with age, with the strongest dependence observed in the radial component. The derived power-law indices are in good agreement with previous observational studies of Galactic disk stars, confirming that WDs preserve the dynamical signatures of long-term disk heating. Moreover, the comparison of $R_{\rm b}$, $R_{\rm g}$, and $R_{\rm gc}$ demonstrates that a significant fraction of the sample has experienced radial migration. The observed increase in radial displacement with stellar age further supports the view that long-term angular momentum redistribution has played a major role in shaping the present-day distribution of local WDs.

Depending on the adopted population-selection criteria, the inferred thin-disk scale height ranges from 123 to 229 pc, whereas the radial scale length remains relatively stable at 3.22-3.49 kpc. The smallest thin-disk scale height is obtained for the orbital selection based on $Z_{\rm max}$, suggesting that orbital parameters provide the most effective separation of thin- and thick-disk populations. In contrast, the thick disk exhibits a substantially larger scale height of approximately 865 pc, consistent with a dynamically older and vertically more extended Galactic component.

The results presented in this study demonstrate that WDs constitute powerful tracers of the dynamical evolution of the Galactic disk. Their well-constrained total ages, combined with precise astrometric data from Gaia, enable investigations of Galactic kinematics, AVR, orbital evolution, and radial migration over timescales inaccessible to most other stellar populations. The consistency of our results with previous studies based on main-sequence stars, red giants, and nearby WDs confirms that the evolutionary pathways inferred for the Galactic disk are largely independent of the stellar tracers employed. 

Future Gaia data releases, together with large-scale spectroscopic surveys such as SDSS-V \citep{Kollmeier2026}, WEAVE \citep{Jin2024}, 4MOST \citep{deJong2019}, DESI \citep{Desi2026}, and MOONS \citep{Cirasuolo2020}, will substantially improve the precision of astrometric, kinematic, and atmospheric parameters, allowing larger and more homogeneous WD samples to be explored \citep[see for example][]{Kilic26a, Kilic26b, Amorim26, Swan2026}. Such advances will facilitate more detailed studies of Galactic dynamical heating, radial migration, and the role of stellar remnants in Galactic archaeology. 

\begin{acknowledgments}
We thank the anonymous referee for their insightful and constructive suggestions that significantly improved the paper. OC thanks the Scientific and Technological Research Council of Türkiye (TÜBİTAK) for funding this research under the 2214-A research project at the University of Oklahoma, as part of his doctoral thesis. MK gratefully acknowledges support by the NSF under grant  AST-2508429, and the NASA under grants 80NSSC24K0380 and 80NSSC24K0436.
\end{acknowledgments}

\vspace{5mm}

\facilities{Gaia \citep{Gaia2016mission, Gaia2023}, SDSS \citep{York2000}}

\software{\texttt{ChatGPT} \citep{ChatGPT}, \texttt{galpy} \citep{Bovy2015}, \texttt{Matplotlib} \citep{Hunter2007}, \texttt{NumPy} \citep{Harris2020},  \texttt{wdwarfdate} \citep{Kiman2022}}

\appendix
\renewcommand{\thetable}{A\arabic{table}}
\setcounter{table}{0}
\section{Catalogue description}
\label{sec:appendix}


\begin{table*}
\centering
\caption{Description of the columns in the catalogue.}
\label{tab:Data}
\scriptsize
\setlength{\tabcolsep}{5pt}
\begin{tabular}{lll}
\hline
\hline
Column & Unit & Description \\
\hline
ID & --- & Order number \\
gaia\_dr3\_source\_id & --- & Gaia DR3 source identifier \\
ruwe & --- & Gaia RUWE \\
S/N & --- & Spectrum SNR \\
binary\_flag & --- & Flag indicating whether the WD is a potential binary \\
$\alpha$ & deg & Right ascension \\
$\delta$ & deg & Declination \\
$l$ & deg & Galactic longitude \\
$b$ & deg & Galactic latitude \\
$d$ & pc & Median geometric distance from Bailer-Jones et al. (2021) \\
$d_{\rm err}$ & pc & Distance uncertainty \\
$\mu_{\alpha}\cos\delta$ & mas\,yr$^{-1}$ & Proper motion in right ascension \\
${(\mu_{\alpha}\cos\delta})_{\rm err}$ & mas\,yr$^{-1}$ & Uncertainty in proper motion in right ascension \\
$\mu_{\delta}$ & mas\,yr$^{-1}$ & Proper motion in declination \\
$(\mu_{\delta})_{\rm err}$ & mas\,yr$^{-1}$ & Uncertainty in proper motion in declination \\
$V_{\rm Rad}$ & km\,s$^{-1}$ & Gravitational-redshift-corrected radial velocity ($V_{\rm grav}$) \\
${(V_{\rm Rad})_{\rm err}}$ & km\,s$^{-1}$ & Uncertainty in radial velocity \\
$X_{\rm gc}$ & pc & Galactocentric Cartesian $X$ coordinate \\
$Y_{\rm gc}$ & pc & Galactocentric Cartesian $Y$ coordinate \\
$Z_{\rm gc}$ & pc & Galactocentric Cartesian $Z$ coordinate \\
$U_{\rm LSR}$ & km\,s$^{-1}$ & Galactic space velocity component $U_{\rm LSR}$ \\
$U_{\rm err}$ & km\,s$^{-1}$ & Uncertainty in $U_{\rm LSR}$ \\
$V_{\rm LSR}$ & km\,s$^{-1}$ & Galactic space velocity component $V_{\rm LSR}$ \\
$V_{\rm err}$ & km\,s$^{-1}$ & Uncertainty in $V_{\rm LSR}$ \\
$W_{\rm LSR}$ & km\,s$^{-1}$ & Galactic space velocity component $W_{\rm LSR}$ \\
$W_{\rm err}$ & km\,s$^{-1}$ & Uncertainty in $W_{\rm LSR}$ \\
$S_{\rm LSR}$ & km\,s$^{-1}$ & Total space velocity relative to the LSR \\
$S_{\rm err}$ & km\,s$^{-1}$ & Uncertainty in the total space velocity \\
$V_{\rm R}$ & km\,s$^{-1}$ & Galactocentric radial velocity component \\
$V_{\rm \phi}$ & km\,s$^{-1}$ & Galactocentric azimuthal velocity component \\
$V_{\rm Z}$ & km\,s$^{-1}$ & Galactocentric vertical velocity component \\
$TD/D$ & --- & Thick-to-thin disk probability ratio \\
$Z_{\rm max}$ & kpc & Maximum distance from the Galactic plane \\
$R_{\rm apo}$ & kpc & Apogalactic distance \\
$R_{\rm peri}$ & kpc & Perigalactic distance \\
e & --- & Orbital eccentricity \\
$P_{\rm orb}$ & Myr & Galactic orbital period \\
$R_{\rm gc}$ & kpc & Present-day Galactocentric distance \\
$R_{\rm g}$ & kpc & Guiding-center radius \\
$R_{\rm b}$ & kpc & Birth radius \\
$T_{\rm eff}$ & K & Photometric effective temperature \\
${(T_{\rm eff})_{\rm err}}$ & K & Uncertainty in photometric effective temperature \\
$\log g$ & cgs & Surface gravity \\
$(\log g)_{\rm err}$ & cgs & Uncertainty in surface gravity \\
$t_{\rm tot}$ & Gyr & Total white dwarf age estimated with \texttt{wdwarfdate} \\
${(t_{\rm tot})_{\rm - err}}$ & Gyr & Lower 1$\sigma$ confidence limit of the total age \\
${(t_{\rm tot})_{\rm + err}}$ & Gyr & Upper 1$\sigma$ confidence limit of the total age \\
$M_{\rm i}$ & $M_\odot$ & Initial (progenitor) mass estimated with \texttt{wdwarfdate} \\
${(M_{\rm i})_{\rm - err}}$ & $M_\odot$ & Lower 1$\sigma$ confidence limit of the initial mass \\
${(M_{\rm i})_{\rm + err}}$ & $M_\odot$ & Upper 1$\sigma$ confidence limit of the initial mass \\
$M_{\rm f}$ & $M_\odot$ & Final (white dwarf) mass estimated with \texttt{wdwarfdate} \\
${(M_{\rm f})_{\rm - err}}$ & $M_\odot$ & Lower 1$\sigma$ confidence limit of the final mass \\
${(M_{\rm f})_{\rm + err}}$ & $M_\odot$ & Upper 1$\sigma$ confidence limit of the final mass \\
\hline
\end{tabular}
\end{table*}

\newpage
\bibliography{references}{}
\bibliographystyle{aasjournalv7}

\end{document}